\documentclass[useAMS,usenatbib]{mn2e}

\usepackage{epsfig} 
\usepackage{multirow}

\renewcommand\({\left(}
\renewcommand\){\right)}
\renewcommand\[{\left[}
\renewcommand\]{\right]}

\newcommand{\ra}{\rightarrow}

\def\lsim{\raise 0.4ex\hbox{$<$}\kern -0.8em\lower 0.62
ex\hbox{$\sim$}}

\def\gsim{\raise 0.4ex\hbox{$>$}\kern -0.7em\lower 0.62
ex\hbox{$\sim$}}

\def\lbar{{\hbox{$\lambda$}\kern -0.7em\raise 0.6ex
\hbox{$-$}}}

\newcommand\eq[1]{eq.~(\ref{#1})}
\newcommand\eqs[2]{eqs.~(\ref{#1}) and (\ref{#2})}
\newcommand\Eq[1]{Equation~(\ref{#1})}
\newcommand\Eqs[2]{Equations~(\ref{#1}) and (\ref{#2})}

\newcommand\eqst[2]{eqs.~(\ref{#1})--(\ref{#2})}
\newcommand\pa{\partial}
\newcommand\p{\partial}

\newcommand\ee{\end{equation}}
\newcommand\be{\begin{equation}}
\def\bea{\begin{array}}
\def\eea{\end{array}}\def\ea{\end{array}}
\newcommand\ees{\end{eqnarray}}
\newcommand\bees{\begin{eqnarray}}

\def\p1{{\bf p}_1}
\def\p2{{\bf p}_2}
\def\k1{{\bf k}_1}
\def\k2{{\bf k}_2}

\newcommand{\dddM}{\kern 0.2em \raise 1.9ex\hbox{$...$}\kern -1.0em \hbox{$M$}}
\newcommand{\dddQ}{\kern 0.2em \raise 1.9ex\hbox{$...$}\kern -1.0em \hbox{$Q$}}
\newcommand{\dddI}{\kern 0.2em \raise 1.9ex\hbox{$...$}\kern -1.0em\hbox{$I$}}
\newcommand{\dddJ}{\kern 0.2em \raise 1.9ex\hbox{$...$}\kern-1.0em
\hbox{$J$}}
\newcommand{\dddcalJ}{\kern 0.2em \raise 1.9ex\hbox{$...$}\kern-1.0em
\hbox{${\cal J}$}}

\newcommand{\dddO}{\kern 0.2em \raise 1.9ex\hbox{$...$}\kern -1.0em
\hbox{${\cal O}$}}
\def\dddz{\raise 1.5ex\hbox{$...$}\kern -0.8em \hbox{$z$}}
\def\dddd{\raise 1.8ex\hbox{$...$}\kern -0.8em \hbox{$d$}}
\def\dddbd{\raise 1.8ex\hbox{$...$}\kern -0.8em \hbox{${\bf d}$}}
\def\ddbd{\raise 1.8ex\hbox{$..$}\kern -0.8em \hbox{${\bf d}$}}
\def\dddx{\raise 1.6ex\hbox{$...$}\kern -0.8em \hbox{$x$}}

\newcommand{\msun}{M_{\odot}}

\def\D{\Delta}
\def\p{\partial}

\def\nn{\nonumber}

\def\s{\sigma}
\def\g{\gamma}

\def\d{\delta}

\def\eps{\epsilon}

\def\dslash{\hspace{-1mm}\not{\hbox{\kern-2pt $\partial$}}}
\def\Dslash{\not{\hbox{\kern-4pt $D$}}}
\def\pslash{\not{\hbox{\kern-2.1pt $p$}}}
\def\kslash{\not{\hbox{\kern-2.3pt $k$}}}
\def\qslash{\not{\hbox{\kern-2.3pt $q$}}}

\newcommand{\inT}{\int_{-\infty}^{\infty}}

\newcommand{\Dl}{\int{\cal D}\lambda}

\newcommand{\Bnp}{B_n^{(p)}}
\newcommand{\Bnq}{B_n^{(q)}}

\title[Conditional probabilities in the excursion set theory]
{Conditional Probabilities in the Excursion Set Theory. 
Generic Barriers and  non-Gaussian Initial Conditions}

\author[Andrea De Simone, Michele Maggiore and  Antonio Riotto]
{Andrea De Simone $^\textrm{\footnotesize{1}}$, Michele Maggiore $^\textrm{\footnotesize{2}}$ and  Antonio Riotto $^\textrm{\footnotesize{3,4}}$\\
\\
$^1$ Institut de Th\'eorie des Ph\'enom\`enes Physiques,
 \'Ecole Polytechnique F\'ed\'erale de Lausanne, CH-1015 Lausanne,
Switzerland\\
$^2$ D\'epartement de Physique Th\'eorique, 
Universit\'e de Gen\`eve, 24 quai Ansermet, CH-1211 Gen\`eve, Switzerland\\
$^3$ CERN, PH-TH Division, CH-1211, Gen\`eve 23,  Switzerland\\
$^4$ INFN, Sezione di Padova, Via Marzolo 8,
I-35131 Padua, Italy
}

\begin{document} 

\date{CERN-PH-TH/2010-323 }    
\pagerange{ 
\pageref{firstpage}-- 
\pageref{lastpage}} 

\maketitle

\label{firstpage} 
\begin{abstract}
The excursion set theory, where density perturbations evolve stochastically with the smoothing scale,
provides a method for computing the dark matter halo mass function.
The computation of the mass function is mapped into   the so-called first-passage time problem in the presence of a moving barrier. The excursion set theory is also a powerful formalism to study other properties of dark matter halos such as   halo bias, accretion 
rate, formation time,  merging rate and the formation history of halos. This is achieved by computing  conditional probabilities
with non-trivial initial conditions, and the conditional two-barrier
first-crossing rate. 
In this paper we  use the  path integral formulation of the  excursion set theory to calculate  analytically these conditional probabilities  
in the presence of a generic moving  barrier,  including the one describing  the ellipsoidal collapse, 
and  for  both   Gaussian and non-Gaussian initial conditions.  While  most of our  analysis 
associated with Gaussian initial 
conditions assumes Markovianity (top-hat in momentum space smoothing, rather than generic filters), 
the non-Markovianity  of the random walks induced by non-Gaussianity is 
consistently accounted for. 
We  compute, for a generic barrier, the first two scale-independent halo bias parameters, the conditional mass function and the
halo formation time probability, including the effects of non-Gaussianities.
We also provide the expression for the two-constant-barrier first-crossing rate
 when non-Markovian effects are induced by a top-hat filter function in
 real space.
\end{abstract}

\section{Introduction}

The distribution in mass of dark matter halos, as well as their clustering properties, formation history, and merging rate, play an important role in many problems
of modern cosmology,  because of their  relevance to the formation and evolution of  galaxies and clusters, and  of their sensitivity to the statistical properties of the primordial density field.
In particular, the most massive halos evolved from rare fluctuations in the primordial density field, so their abundance and  clustering properties
are sensitive probes of primordial 
non-Gaussianities \citep{MLB,GW,LMV,MMLM,KOYAMA,MVJ,RB,RGS,LV,MR3,lam,porciani},
which could be detected or significantly constrained by
various planned large-scale galaxy surveys, see,  e.g.  \cite{Dalal} and \cite{CVM}. 
Furthermore, the primordial non-Gaussianities (NG)
alters the clustering of dark matter halos inducing a scale-dependent
bias on large 
scales \citep{Dalal,MV,slosar,tolley}  while even for small primordial
NG the evolution of perturbations on super-Hubble scales yields
extra contributions on 
smaller scales \citep{bartolosig,MV2009,bartolo2010}.

The halo mass function can be written as
\be
\label{dndMdef}
\frac{dn(M)}{dM} = f(\s) \frac{\bar{\rho}}{M^2} 
\frac{d\ln\s^{-1} (M)}{d\ln M}\, ,
\label{massfunction}
\ee
where $n(M)$ is the  number density of dark matter halos of mass $M$,
$\s(M)$ is the variance of the linear density field smoothed on a
scale $R$ corresponding to a mass $M$, and
$\bar{\rho}$ is the average density of the universe. 
The basic problem is therefore the computation of the function $f(\s)$.
Analytical computations of the halo mass function 
are typically based  on
Press-Schechter (PS) 
theory \citep{PS} and its extension~\citep{PH90,Bond} known as excursion set
theory	(see \cite{Zentner} for a  review).
In  excursion set theory the density
perturbation depends stochastically 
with the smoothing scale,
and the problem of computing the probability of halo formation is
mapped into the so-called first-passage time problem in the presence
of a barrier. With standard manipulations (see e.g.~\cite{Zentner}), the function $f(\sigma)$
which appears in (\ref{massfunction}) is related to the first-crossing rate ${\cal F}$ by
$f(\sigma)=2\sigma^2 {\cal F}(\sigma^2)$.

In  a recent series of papers \citep{MR1,MR2,MR3}  (hereafter MR1, MR2 and MR3, respectively), the original formulation of excursion set theory has been extended to deal analytically with 
the non-Markovian effects which are induced either by  the use of  a realistic filter function, or by   
non-Gaussianities in the primordial density field. In the  original formulation of \cite{Bond} the problem with the density field smoothed using a top-hat filter in wavenumber space was solved analytically and numerical techniques were adopted for   the case of non-Markovian noises. The use of a top-hat window function in momentum space has the technical advantage that the evolution of the smoothed density field with the smoothing scale becomes Markovian,
but its important drawback is that is it is not possible to associate a well-defined  mass to a region smoothed with such a filter (see \cite{Bond,Zentner,MR1}). For any other choice of filter function such as 
a top-hat function in real space (for which the relation between the mass $M$ and the smoothing scale $R$ is well-defined and is simply  $M=(4/3)\pi R^3\bar{\rho}$) the actual evolution of the smoothed density field with $R$ is non-Markovian. 	

The same happens  if the initial conditions for the gravitational potential and/or the density contrast are non-Gaussian and the problem  was solved first in MR3.
The basic idea is to reformulate the first-passage time problem in the presence of a barrier in terms of the computation of a path integral with a boundary (i.e. over a sum over all ``trajectories'' $\d(S)$ that always stay below the barrier), and then to use  standard
results  from quantum field theory and statistical mechanics to express  this path integral in terms of the connected correlators of the theory. 
This  allows us to  include
the effect of non-Markovianities arising, e.g., from the  non-Gaussianities. In particular, in 
MR3 we have shown how to include the effect of a non-vanishing bispectrum, while the case of a non-vanishing trispectrum was considered in \cite{MR4} (see also \cite{D'Amico:2010ta} for an approach to non-Gaussianities which combines our technique with the saddle  point method developed in
\cite{MVJ}).

An essential ingredient of excursion set theory is a model for the collapse of a dark matter halo. In its simplest implementation, one uses the spherical collapse model. This
model, however, is certainly a significant  over-simplification of the complicated dynamics leading to halo formation and
can be improved in different, complementary, ways.  A crucial step was taken by  \cite{SMT} who took into account the fact that actual halos are triaxial \citep{BBKS,BondMyers} and showed that 
an ellipsoidal collapse  model can be implemented, within the excursion set theory framework, by computing the first-crossing rate in the presence of a barrier $B_{\rm ST}(S)$,
\be\label{B(S)}\label{ST}
B_{\rm ST}(S)\simeq \sqrt{a}\delta_c(z)\[ 1 +0.4\(\frac{S}{a\delta_c^2(z)}\)^{0.6}\]\, ,
\ee
which depends on $S\equiv \sigma^2$ (``moving barrier"), rather than taking the   value $\d_c(z)$ of the spherical collapse,
which is redshift dependent, but independent of $S$.
Physically this reflects the fact  that low-mass halos
(which corresponds to  large $S$) have larger deviations from
sphericity and significant shear, that opposes  collapse. 

Notice that, to improve the agreement between the prediction from the
excursion set theory with an ellipsoidal collapse and the N-body 
simulations, \cite{SMT} also found that
it was necessary to multiply $\d_c(z)$ by $\sqrt{a}$, where  $\sqrt{a}\simeq 0.84$ was obtained
by requiring that their mass function 
fits  the GIF simulation. 
In MR2 we proposed a physical justification for the introduction of this parameter in the halo mass function, suggesting 
that some of the physical complications inherent to a realistic description of halo formation could be included in the excursion set theory framework, at least at an effective level, by treating the critical threshold for collapse  as a stochastic variable, whose scatter reflects a number of complicated aspects 
of the underlying dynamics (see also \cite{Audit,LeeS,SMT} for earlier related ideas). Solving the first-passage time problem in the presence of a barrier which is diffusing around 
its mean value, 
it was found in MR2 that the coefficient $a$ can be related to the diffusion coefficient $D_B$  of the
stochastic barrier as
$a=1/(1+D_B)$.  The numerical value of $D_B$, and therefore the corresponding value of $a$, depends among other things on the algorithm used for identifying halos. From recent N-body simulations that studied the properties of the collapse barrier, a value $D_B\simeq 0.25$ was deduced in MR2,   predicting   $a\simeq 0.80$, in  agreement within the accuracy of the MR1 prediction ($\sim$ 20\%) with the value of $a$ extracted directly from a fit to the mass function 
(see also \cite{cor} for recent related work).

The path-integral formulation developed in MR1 and MR3 was restricted to the case of a constant barrier $\d_c(z)$ and it was subsequently generalized to the case of the ellipsoidal moving barrier 
in \cite{DMR}.
In the present paper we  further develop  the path-integral formulation of  the excursion set theory  to calculate, for a generic moving barrier and for Gaussian and non-Gaussian initial conditions, other basic quantities  necessary to characterize the physics
of dark matter halos like halo bias, accretion rates, formation times, merging,  halo assembly bias and so on. 

We know that 
dark matter halos typically form at sites of high density peaks.  The
spatial distribution of dark matter halos is therefore a biased tracer of
the underlying mass distribution.  A standard way to quantify this
difference between halos and mass is to use a bias parameter $b_h$, which
can be defined as the ratio of the overdensity of halos to mass, or as the
square root of the ratio of the two-point correlation function (or power
spectrum) of halos to mass. 
Like the halo mass function, analytic expressions for the halo bias can be
obtained from the excursion set theory based on the spherical gravitational
collapse model \citep{CK89,Bond,MW96} and for the ellipsiodal one \citep{SMT}.
The approach to the clustering evolution is based on a generalization of the so-called peak-background split (\cite{BBKS}) which basically consists in splitting the mass perturbations in a fine-grained (peak) component  filtered on a scale $R$
and a
coarse-grained (background) component  filtered on a scale $R_0\gg R$. The underlying idea is to ascribe the collapse of objects on small scales to the high frequency modes of the density fields, while the action of large-scale structures of these non-linear condensations is due to a shift of the local background density. In  the excursion set theory  the problem of computing the probability of halo formation is
mapped into the  first-passage time problem of a random walk which starts from a given value
of the density contrast $\delta_0$ at a given radius $R_0$ corresponding to a given value of the variance  $\sigma(M_0)$. When the random walk performed by the smoothed density contrast is Markovian, the first-crossing rate is easily computed by a simple shift of the initial conditions. 
This is due to the fact that, being the noise white,  the memory about the way the system arrived at the point $\delta_0$ at a given time is lost. 
On the contrary, when the random walk is non-Markovian, the system has memory effects and it remembers how it arrived
at $\delta_0$. This influences the subsequent first-crossing rate. The computation of the
halo bias mass function in the case in which the non-Markovianity is induced by the choice of
a top-hat window function in real space, and within a spherical collapse model, has been recently performed in \cite{ma}. In this paper we
perform the calculation of the halo bias parameters for the ellipsoidal barrier and when non-Gaussian initial conditions introduce non-Markovianity, see also \cite{porciani} for a treatment of NG halo bias.

The excursion set theory is also a powerful formalism for studying the formation history  of halos.
The most immediate quantity of interest is the conditional mass function. Given a halo of
mass $M_0$ at redshift $z_a$, one can compute the average manner in which this mass was partitioned
among smaller halos at some  higher redshift $z_b>z_a$. The conditional mass function is simply the average number of halos of mass $M_n$ at redshift $z_b$ that are incorporated
into an object of mass $M_0$ at redshift $z_a$. 
In the language of excursion set theory this  can be formulated as 
a two-barrier problem, i.e. in terms of the conditional first crossing rate,
${\cal F}(B_b(S_n), S_n|B_a(S_0), S_0)$, describing the rate at which
trajectories make their first crossing of the barrier $B_b(S)\equiv B(S,z=z_b)$ at a value $S=S_n$,  corresponding to the mass $M_n$, under the condition that, at an
earlier ``time" $S=S_0$ (corresponding to the mass $M_0$; recall that decreasing the variance $S$  the corresponding mass $M(S)$ increases, so $S_0<S_n$ means $M_0>M_n$), they crossed the threshold $B_a(S)\equiv B(S,z=z_a)$.
Then, a halo of mass
$M_0$ has its mass partitioned on average among a spectrum of halos at redshift $z_b$
as~\citep{lc,Zentner}
\be
\label{condmassfunction}
\frac{{\rm d}n(M_n|M_0)}{{\rm d}M_n}=\frac{M_0}{M_n}\, 
{\cal F}(B_b(S_n),S_n|B_a(S_0),S_0)\left|\frac{{\rm d}S_n}{{\rm d} M_n}\right|\, . 
\ee
The function
${\cal F}(S_n,B_b(S_n)|S_0,B_a(S_0))$ gives the probability of the second barrier first-crossing at a 
particular value of $S_n$, while the factor $(M_0/M_n)$ converts it from a probability per unit mass
of halo $M_0$ into the number of halos of mass $M_n$. The two-barrier result can also be manipulated 
to yield the average mass accretion rate, halo formation time and so on.

The relationship between the unconditional mass function and the
first-crossing distribution associated with barrier-crossing random
walks has been extended to obtain the conditional mass function
of halos  by \cite{Bond} and \cite{lc}  within the spherical collapse  (the  so-called
extended Press-Schechter  model).
The  two-barrier first-rate probability  has a simple analytic form in the
constant barrier spherical collapse model. Again, this is  because the random
walk performed by the smoothed density contrast is Markovian.
For a moving barrier (such as the ellipsoidal collapse model), however,
exact analytic forms have been found only for the special case
of a linear barrier \citep{Sheth1998} while \cite{lam}
 have proposed a better motivated approach to the problem by offering a Taylor series-like approximation
for a general moving barrier, see also \cite{Giocoli2007}.  \cite{zhang} have provided  analytical expressions for the 
two-barrier first-crossing rate for the ellipsoidal collapse and Gaussian initial conditions.
In this paper we compute this conditional probability using the path-integral formulation for a generic 
moving barrier for Gaussian and   non-Gaussian initial conditions. A by-product of such a calculation
is the determination of the halo formation time probability.

The paper is organized as follows. In section \ref{sect:pi}  we summarize the basic ingredients for the calculation of the first-crossing rate from the excursion set theory and a generic moving barrier. In section \ref{sect:bias} we compute the conditional probability necessary to deduce the  halo bias parameters, both in the Gaussian and non-Gaussian case. Section \ref{sect:2barrier} contains the computation of the two-barrier
first-crossing rate for a generic moving barrier and again for both Gaussian and non-Gaussian
initial conditions. In section \ref{sect:formation} we present our results for the halo formation time 
probability. Finally, Section \ref{sect:conclusions}
contains our conclusions and a summary of the main results, 
while some technical material is collected in the Appendices. In particular, Appendix \ref{app:fits} contains some useful numerical fits, while Appendix \ref{app:realspace} contains the computation of the
two-barrier first crossing rate including the non-Markovian effects coming from the choice of a top-hat filter in real space.

\section{Path integral formulation of  excursion set theory for a moving barrier}
\label{sect:pi}
Let us discuss the basic points of the original formulation of 
excursion set theory for a moving barrier. We will closely follow  MR1 and \cite{DMR}; at the expense of being
ripetitive, we will report here various details that the reader can find in these references. This will hopefully help
to follow and speed up the calculations of the subsequent sections. 

In the excursion set theory, one
considers the density field $\d$  smoothed over a radius $R$, and studies its stochastic evolution as a function
of the smoothing scale $R$.
As it was found in the classical paper by
\cite{Bond},
when the density $\d(R)$  is smoothed with a sharp filter in wavenumber
space, and the density fluctuations have Gaussian statistics, the smoothed
density field satisfies the equation
\be\label{Langevin1}
\frac{\pa\d(S)}{\pa S} = \eta(S)\, ,
\ee
where $S=\s^2(R)$ is the variance of the linear density field
smoothed on the scale $R$ and computed with a sharp filter in wavenumber
space, while
$\eta (S)$ is a stochastic variable that satisfies 
\be\label{Langevin2}
\langle \eta(S_1)\eta(S_2)\rangle =\d_D (S_1-S_2)\, ,
\ee
where $\d_D$ denotes the Dirac delta function.
\Eqs{Langevin1} {Langevin2} are   the same as a Langevin equation with a
Dirac-delta  noise $\eta(S)$, with the variance $S$ formally playing the role
of time. Let us denote by $\Pi(\d,S)d\d$ the probability density that 
the variable $\d(S)$ reaches a value between $\d$ and $\d+d\d$ by ``time" $S$. 
In the general non-Markovian case it is not possible to derive  a simple, local,
differential equation for $\Pi(\d,S)$ (indeed, it can be shown that
$\Pi(\d,S)$ rather satisfies a complicated integro-differential equation which is non-local with respect to ``time" $S$, see eq.~(83) of MR1), so one cannot proceed as in the Markovian case where, as we will review below,  
$\Pi(\d,S)$ is determined by the solution of the Fokker-Planck equation with appropriate boundary conditions. Rather, we
construct the probability distribution $\Pi(\d,S)$ directly by summing over all
paths that never exceeded the corresponding threshold, i.e. 
by writing $\Pi(\d,S)$ as a
path integral with a boundary.	To obtain such a representation, 
we consider an ensemble of
trajectories all starting at $S_0=0$ from an initial position
$\d(0) =\d_0$ 
and we follow them for a ``time'' $S$.	
We discretize the interval $[0,S]$ in steps
$\D S=\eps$, so $S_k=k\eps$ with $k=1,\ldots n$, and $S_n\equiv S$. 
A trajectory is  then defined by
the collection of values $\{\d_1,\ldots ,\d_n\}$, such that $\d(S_k)=\d_k$
and  $B(S_i)=B_i$.
The probability density in the space of  trajectories is 
\be\label{defW}
W(\d_0;\d_1,\ldots ,\d_n;S_n)\equiv \langle
\d_D (\d(S_1)-\d_1)\ldots \d_D (\d(S_n)-\d_n)\rangle\, ,
\ee
where  $\d_D$
denotes the Dirac delta. Then the probability of arriving in $\d_n$ in a
``time'' $S_n$, starting from an initial value $\d_0$, without ever
going above the threshold, is (\cite{Bond})
\begin{eqnarray}
\Pi_{\rm mb} (\d_n;S_n)
& \equiv&\int_{-\infty}^{B_1} d\d_1\ldots \int_{-\infty}^{B_{n-1}}d\d_{n-1}\, 
\\
&&\times W(\d_0;\d_1,\ldots ,\d_{n-1},\d_n;S_n).\nn
\ees
The label ``${\rm mb}$''  in $\Pi_{\rm mb}$ stands for moving barrier.
The function  $W(\d_0;\d_1,\ldots ,\d_{n-1},\d_n;S_n)$ can
be expressed in terms of the connected correlators of the theory,
\be
W(\d_0;\d_1,\ldots ,\d_n;S_n)=\Dl \, {\rm e}^Z\, ,
\ee
where
\be
\Dl \equiv
\inT\frac{d\lambda_1}{2\pi}\ldots\frac{d\lambda_n}{2\pi}\, ,
\ee
and
\bees
Z&=& i\sum_{i=1}^n\lambda_i\d_i\label{defZ}\\
&&+\sum_{p=2}^{\infty}
\frac{(-i)^p}{p!}\,
 \sum_{i_1=1}^n\ldots \sum_{i_p=1}^n
\lambda_{i_1}\ldots\lambda_{i_p}\,
\langle \d_{i_1}\ldots\d_{i_p}\rangle_c\, .\nn
\ees
Here $\langle \d_{1}\ldots\d_{n}\rangle_c$ denotes the connected $n$-point
correlator. So
\be\label{Piexplicit}
\Pi_{\rm mb}(\d_0;\d_n;S_n)=\int_{-\infty}^{B_1} d\d_1\ldots \int_{-\infty}^{B_{n-1}}d\d_{n-1}\,
\Dl\, {\rm e}^Z\, .
\ee

When $\d(S)$ satisfies \eqs {Langevin1} {Langevin2} (which is the case
for sharp filter in wavenumber space) the two-point function can be easily
computed, and is given by
\be\label{twopoint}
\langle\delta(S_i)\delta(S_j)\rangle = {\rm min}(S_i,S_j)\, .
\ee
In the rest of this section we will restrict ourselves to the Gaussian and Markovian  case.

\subsection{The case of Markovian noise}
\noindent
Taking the derivative with respect to the time $S_n\equiv S$ of 
eq.~(\ref{Piexplicit}) and using the 
fact that, when multiplying $\exp\{i\sum_i\lambda_i\delta_i\}$, 
 $i\lambda_j$ $(j=1,\cdots, n)$ can be replaced $\partial_j\equiv \pa/\pa \delta^j$, we discover that $\Pi_{\rm mb} (\d_n;S_n)$ satisfies the Fokker-Planck (FP) equation
\be\label{FP35}
\frac{\partial \Pi_{\rm mb} (\d_n;S_n)}{\partial S_n}=\frac{1}{2}\frac{\partial^2\Pi_{\rm mb} (\d_n;S_n)}{\partial\delta^2}\, .
\ee
In the continuum limit, the boundary condition to be imposed on the solution of \eq{FP35} is 
\be\label{Boundcond}
\Pi_{\rm mb}(\d_n ;S_n)=0 \qquad \textrm{for}\quad \delta_n\geq B_n\,.
\ee
In the continuum limit the first-crossing rate is then given by
\begin{eqnarray}
{\cal F}_{\rm mb}(S_n)  &\hspace*{-2mm}=&\hspace*{-2mm} -\frac{\pa}{\pa S}
\int_{-\infty}^{B_n}d\d_n\, \Pi_{\rm mb}(\d_n;S_n)\\
&\hspace*{-2mm}=&\hspace*{-2mm}-{dB_n\over dS_n}\Pi_{\rm mb}(B_n,S_n)-\int_{-\infty}^{B_n}d\d_n\, \frac{\partial\Pi_{\rm mb}(\d_n;S_n)}{\partial S_n}\, .\nonumber
\ees
The first term on the right-hand side vanishes because of the boundary condition, while the second term can be written in  a more convenient form using the FP equation (\ref{FP35}),
so
\bees
{\cal F}_{\rm mb}(S_n)&=&
-\frac{1}{2}\int_{-\infty}^{B_n}d\d\, \frac{\partial^2\Pi_{\rm mb}(\d_n;S_n)}{\partial \d^2}\nn\\
&=&-\frac{1}{2}\left.\frac{\partial\Pi_{\rm mb}(\d_n;S_n)}{\partial \d_n}\right|_{\delta=B_n}\, .
\label{firstcrossmoving}
\end{eqnarray}
To compute the probability
$\Pi_{\rm mb}(\delta_n,S_n)$ we proceed in the following way. At every $i$-th step of the path integral we Taylor expand the barrier around its final value, 
as first suggested by \cite{lam},
\be
B_i=B_n+\sum_{p=1}^\infty\frac{\Bnp}{p!}\, \left(S_i-S_n\right)^p\, ,
\ee
where
\be
\Bnp\equiv  \frac{d^p B(S_n)}{d S_n^p}\, , 
\ee
(so in particular $B_n^{(0)}=B(S_n)$).
We now perform a shift in the integration variables $\delta_i$ ($i=1,\ldots,n-1$) in the path integral
\be
\label{shift}
\delta_i\rightarrow \delta_i-\sum_{p=1}^\infty\frac{\Bnp}{p!}
\, \left(S_i-S_n\right)^p\, .
\ee
Then $\Pi_{\rm mb} (\d_n;S_n)$ can be written as 
\be
\Pi_{\rm mb} (\d_n;S_n)=
 \int_{-\infty}^{B_n} d\d_1\ldots \int_{-\infty}^{B_n}d\d_{n-1}\,\Dl \,\,{\rm e}^Z
 \ee
 where
 \bees
 Z&=&i\sum_{i=1}^n\lambda_i\d_i
 -\frac{1}{2}\sum_{i,j=1}^n\lambda_{i}\lambda_{j}\,{\rm min}(S_i,S_j)\nn\\
&&
 +i\sum_{i=1}^{n-1}\lambda_i\sum_{p=1}^\infty\frac{\Bnp}{p!}
\, \left(S_i-S_n\right)^p .
\ees
We next expand 
\begin{eqnarray}
&&\hspace*{-5mm}{\rm exp}\left\{i\sum_{i=1}^{n-1}\lambda_i\sum_{p=1}^\infty\frac{\Bnp}{p!}
\, \left(S_i-S_n\right)^p\right\}\nn\\
&&\hspace*{-5mm}
\simeq 1+i\sum_{i=1}^{n-1}\lambda_i\sum_{p=1}^\infty\frac{\Bnp}{p!}\, \left(S_i-S_n\right)^p
\label{eq22next}\\
&&\hspace*{-5mm}
-\frac{1}{2}\sum_{i,j=1}^{n-1}\lambda_i\lambda_j\sum_{p,q=1}^\infty\frac{\Bnp\Bnq}{p!q!}
\left(S_i-S_n\right)^p
\, \left(S_j-S_n\right)^q+\cdots\, ,\nn
\end{eqnarray}
and we  write $\Pi_{\rm mb} (\d_n;S_n)$ as 
\bees
\Pi_{\rm mb} (\d_n;S_n)&=&\Pi^{(0)}_{\rm mb} (\d_n;S_n)+\Pi^{(1)}_{\rm mb} (\d_n;S_n)\nn\\
&&+\Pi^{(2)}_{\rm mb} (\d_n;S_n)
+\cdots\, .\label{Pi0Pi1Pi2}
\ees
For the zero-th order term $\Pi^{(0)}_{\rm mb}$ we can immediately take the continuum limit, using the results of MR1, and we get the standard  probability density of excursion set theory in the Markovian and Gaussian case,
\begin{eqnarray}
\Pi^{(0)}_{\rm mb} (\d_n;S_n)=\frac{1}{\sqrt{2\pi S_n}}\,
\[  {\rm e}^{-\d_n^2/(2S_n)}- {\rm e}^{-(2B_n-\d_n)^2/(2S_n)} \]\,. 
\hspace{-1cm}\nn\\
\end{eqnarray}
The terms $\Pi^{(1)}_{\rm mb}$ and $\Pi^{(2)}_{\rm mb}$ are given by
\bees
\Pi^{(1)}_{\rm mb} (\d_n;S_n)&=&\sum_{i=1}^{n-1}\int_{-\infty}^{B_n} d\d_1\ldots d\d_{n-1}
\sum_{p=1}^\infty\frac{\Bnp}{p!}
\label{Pi1}\\
&&\times\left(S_i-S_n\right)^p \partial_i W^{\rm gm}(\delta_0;\d_1,\ldots,\delta_n; S_n)\, ,\nn
\ees
and
\begin{eqnarray}
\label{s2}
&&\hspace*{-5mm}\Pi^{(2)}_{\rm mb} (\d_n;S_n)=\frac{1}{2}\sum_{i,j=1}^{n-1}\int_{-\infty}^{B_n} d\d_1\ldots d\d_{n-1}
\sum_{p,q=1}^\infty\frac{\Bnp\Bnq}{p!q!}\nn\\
&&\hspace*{-5mm}\times\left(S_i-S_n\right)^p
\left(S_j-S_n\right)^q
\partial_i\partial_j W^{\rm gm}(\delta_0;\d_1,\ldots,\delta_n; S_n)\, ,
\end{eqnarray}
where 
\be\label{W}
W^{\rm gm}(\d_0;\d_1,\ldots ,\d_n;S_n)=\frac{1}{(2\pi\eps)^{n/2}}\, 
{\rm e}^{-\frac{1}{2\eps}\,
\sum_{i=0}^{n-1}  (\d_{i+1}-\d_i)^2}\hspace{-2mm},
\ee
and  superscript ``gm'' (Gaussian-Markovian) reminds us that this value of
$W$ is computed for Gaussian fluctuations, and when the evolution with respect to the
smoothing scale is Markovian. Their continuum limit is more subtle, and can be computed using the technique developed in MR1, as we review below.

We have therefore formally expanded $\Pi_{\rm mb}(\d_n,S_n)$ in a series of terms
$\Pi^{(1)}_{\rm mb}$, $\Pi^{(2)}_{\rm mb}$, etc., in which each term is itself given by an infinite sum over indices $p, q,\ldots$ . 
We have to evaluate the continuum limit of objects  such as 
\be\label{expr}
\sum_{i=1}^{n-1}F(S_i)\int_{-\infty}^{B_n} 
d\delta_1\ldots d\delta_{n-1}\,\pa_iW^{\rm gm}(\delta_0;\delta_1,\ldots
,\delta_n;S_n)\, ,\nn
\ee
where $F$ denotes a generic function.
To compute this expression we  integrate $\pa_i$ by parts,
\bees
&&\int_{-\infty}^{B_n} 
d\delta_1\ldots d\delta_{n-1}\,\pa_iW^{\rm gm}(\delta_0;\delta_1,\ldots
,\delta_n;S_n)\nn\\
&=&\int_{-\infty}^{B_n} d\delta_1\ldots \widehat{d\d_i}\ldots 
d\delta_{n-1}\label{byparts}\\
&&\times W(\delta_0;\delta_1,\ldots, \delta_i=B_n,\ldots
,\delta_{n-1},\delta_n;S_n)\, ,\nn
\ees
where
the notation $ \widehat{d\d_i}$ means that we must omit $d\delta_i$ from
the list of integration variables. 
We next observe that $W^{\rm gm}$ satisfies
\bees\label{facto}
&&W^{\rm gm}(\delta_0;\delta_1,\ldots, \delta_{i}= B_n, 
\ldots ,\delta_n; S_n)\nn\\
&=& 
W^{\rm gm}(\delta_0;\delta_1,\ldots, \delta_{i-1}, B_n;S_i)\nn\\
&&\times W^{\rm gm}(B_n; \delta_{i+1}, \ldots ,\delta_n; S_n-S_i)\, ,
\ees
as can be verified directly from its explicit expression (\ref{W}). Then
\bees\label{WPiPi}
&&\int_{-\infty}^{B_n}d\delta_1\ldots d\delta_{i-1}
\int_{-\infty}^{B_n}  d\delta_{i+1}\ldots	d\delta_{n-1}\nn\\
&&\times W^{\rm gm}(\delta_0;\delta_1,\ldots, \delta_{i-1}, B_n;S_i)\nn\\
&&\times
W^{\rm gm}(B_n; \delta_{i+1}, \ldots ,\delta_n; S_n-S_i)\nn\\
&&= \Pi^{\rm gm}(\delta_0;B_n;S_i)
\Pi^{\rm gm}(B_n;\delta_n;S_n-S_i)\, ,  
\ees
and to compute the expression given in \eq{expr} we must compute objects such as
\be\label{Pimem1}
\sum_{i=1}^{n-1}F(S_i) \Pi^{\rm gm}(\delta_0;B_n;S_i)
\Pi^{\rm gm}(B_n;\delta_n;S_n-S_i) .
\ee
We then need to know $\Pi^{\rm gm}(\delta_0;B_n;S_i)$. By definition, in the continuum limit this quantity
vanishes, since its second argument is equal to the the threshold value $B_n$. However, in the continuum limit the sum over $i$ becomes
$1/\eps$ times an integral over an intermediate time variable $S_i$,
\be\label{sumint}
\sum_{i=1}^{n-1}\ra \frac{1}{\eps}\int_o^{S_n} dS_i\, ,
\ee
so we need to know how $\Pi^{\rm gm}(\delta_0;B_n;S_i)$ approaches
zero when $\eps\ra 0$.	In MR1 it was proven that it vanishes as
$\sqrt{\eps}$, and that
\begin{eqnarray}\label{Pigammafinal}
&&\hspace*{-5mm}\Pi^{\rm gm} (\delta_0;B_n;S_n)=
\sqrt{\eps}\,  \,
\frac{B_n-\delta_0}{\sqrt{\pi}\, S_n^{3/2}} {\rm e}^{-(B_n-\delta_0)^2/(2S_n)} +{\cal
O}(\eps) .\nonumber\\
&&
\ees
Similarly, for $\delta_n<B_n$,
\begin{eqnarray}\label{Pigammafinalbis}
&&\hspace*{-5mm}\Pi^{\rm gm} (B_n;\delta_n;S_n)=
\sqrt{\eps}\, \, 
\frac{B_n-\delta_n}{\sqrt{\pi}\, S_n^{3/2}} {\rm e}^{-(B_n-\delta_n)^2/(2S_n)} +{\cal
O}(\eps).\nonumber\\
&&
\ees
In the following, we will also need the expression for $\Pi^{\rm gm}$ 
with the  first and second 
argument both equal to $B_n$, which is given by (see again MR1)
\be\label{Pigammafinalter}
\Pi^{\rm gm}(B_n;B_n;S)=
{\eps\over \sqrt{2\pi} S_n^{3/2}}\, .
\ee 
In order to finalize the computation, we must either perform some approximation, or identify a suitable small parameter, and organize 
the terms in a systematic expansion in such a small parameter. In 
\cite{DMR} we have discussed in detail two  different expansion techniques 
(one based on a systematic expansion in derivatives for a slowly varying barrier, and the other in which a large number of terms are resummed), which were
shown to provide  very close  numerical results. Furthermore,
it was found that the results obtained with these systematic expansions are in the end numerically very close to that obtained with a simpler albeit more empirical procedure,  which amounts to
approximating  $(S_n-S_i)^{p-1}\simeq S_n^{p-1}$ inside the integrals in \eqs{Pi1}{s2},  and at the same time truncating the sum over $p$ in \eq{Pi1} to $p=5$ (while, in this approximation, $\Pi_{\rm mb}^{(2)}$ does not contribute).
This is in fact equivalent to the approximation made  by \cite{ST2002} (ST in the following) and gives the same results. However, as discussed in Section~3.1 of \cite{DMR}, when one makes the approximation $(S_n-S_i)^{p-1}\simeq S_n^{p-1}$,  one must also necessarily truncate the sum to a maximum value, otherwise the first-crossing rate  resums to a trivial result, where all corrections due to the ellipsoidal barrier disappear. Therefore, the procedure of replacing
$(S_n-S_i)^{p-1}\ra S_n^{p-1}$ inside the integrals and, {\em at the same time}, truncating the sum, must be viewed as a simple heuristic procedure to get a result which is numerically close to the result of more systematic expansions. Since this procedure is technically much simpler than the systematic expansions discussed in \cite{DMR}, and works well numerically, we will adopt it in the following.

We first compute $\Pi_{\rm mb}^{(1)}$.  Before performing the above approximation, the  
 expression of  $\Pi^{(1)}_{\rm mb} (\d_n;S_n)$ in eq.~(\ref{Pi1}) can be rewritten as
\begin{eqnarray}
\Pi^{(1)}_{\rm mb} (\d_n;S_n)&=&\frac{B_n(B_n-\delta_n)}{\pi}\sum_{p=1}^{\infty}\frac{(-1)^p}{p!}\Bnp\nn
\\
&&\times\int_0^{S_n}d S_i\frac{\left(S_n-S_i\right)^{p-(3/2)}
}{S_i^{3/2}}\nn\\
&&\times {\rm e}^{-B_n^2/(2S_i)}{\rm e}^{-(B_n-\delta_n)^2/[2 (S_n-S_i)]}\, .
\label{ainApp}
\end{eqnarray}
Since this  integral  is finite in the limit $\delta_n\rightarrow B_n$,  taking  the approximation
$(S_n-S_i)^{p-1}\simeq (S_n)^{p-1}$ does not alter the convergence properties of the integral, but simplifies significantly its computation, since
\begin{eqnarray}
&&\int_0^{S_n}d S_i\frac{1}{S_i^{3/2}(S_n-S_i)^{1/2}}\nn\\
&&\times {\rm e}^{-B_n^2/(2S_i)} {\rm e}^{-(B_n-\delta_n)^2/(2 (S_n-S_i))}
\nonumber\\
&=&\frac{\sqrt{2\pi}}{B_n}\frac{1}{S_n^{1/2}}{\rm exp}\left\{-\frac{(2B_n-\delta_n)^2}{2 S_n}\right\}\, ,
\label{eq108}
\end{eqnarray}
so in this approximation  $\Pi^{(1)}_{\rm mb} (\d_n;S_n)$ is given by
\begin{eqnarray}
\label{zz}
\Pi^{(1, {\rm ST})}_{\rm mb} (\d_n;S_n)&=&
{2(B_n-\delta_n)\over \sqrt{2\pi} S_n^{3/2}}
 {\rm e}^{-(2B_n
 -\delta_n)^2/(2 S_n)}\nn\\
&&\times \sum_{p=1}^5\frac{(-S_n)^p}{p!}B_n^{(p)}\, .
\end{eqnarray}
where the superscript ``ST" reminds us that
we have performed the approximations that are equivalent to those which give the ST mass function.
A reason why this approximation works well is that (at least for what concerns $\Pi^{(1)}_{\rm mb}$) the terms which are neglected give contributions proportional to  higher powers of
$(B_n-\delta_n)$. Since in the end the mass function is obtaned
from  the first-crossing rate
(\ref{firstcrossmoving}),  we actually only need the first derivative of $\Pi_{\rm mb} (\d_n;S_n)$ evaluated
at $\delta_n=B_n$, and terms proportional to $(B_n-\delta_n)^{N}$ with $N\geq 2$ give a vanishing contribution.

Higher-order contributions to the first-crossing rate vanish. In fact, in the same approximation one finds \citep{DMR} that
$\Pi^{(n,{\rm ST})}_{\rm mb}$ vanishes as $(B_n-\d_n)^n$ for $\d_n\ra B_n$, so
its first derivative $\pa\Pi^{(n),{\rm ST}}_{\rm mb}/\pa\d_n$ evaluated in $\d_n=B_n$,
which according to \eq{firstcrossmoving}
gives its contribution to the first-crossing rate, vanishes for all $n\geq 2$.

The total first-crossing rate for a moving barrier, in the approximation discussed above, 
is therefore given by 
\begin{eqnarray}
\label{aaprime}
{\cal F}_{\rm mb}(S_n)&=&\frac{{\rm e}^{-B_n^2/(2S_n)}}{\sqrt{2\pi}S_n^{3/2}}
\sum_{p=0}^5\frac{(-S_n)^p}{p!} \frac{\partial^p B_n}{\partial S_n^p}\nonumber\\
&=&\frac{{\rm e}^{-B_n^2/(2S_n)}}{\sqrt{2\pi}S_n^{3/2}}\left(B_n+{\cal P}(S_n)\right)\, ,
\end{eqnarray}
where

\be
{\cal P}(S_n)\equiv{\cal P}_n=\sum_{p=1}^5\frac{(-S_n)^p}{p!} \frac{\partial^p B_n}{\partial S_n^p}\, .
\ee
 When applied to the ellipsoidal barrier given in
eq.~(\ref{ST}), one recovers the ellipsoidal collapse result of 
\cite{ST2002}.
\begin{eqnarray}
{\cal F}_{\rm ST}(S_n)&\simeq& 
\frac{\sqrt{a}\,\delta_c(z)}{\sqrt{2\pi }S_n^{3/2}}{\rm e}^{-B_n^2/(2S_n)}\Bigg[1+ \nn\\
&&\left. +0.4\sum_{p=0}^5(-1)^p{0.6\choose p}
\left(\frac{S_n}{a \delta_c^2(z)}\right)^{0.6}\right]\nn\\
&&\hspace*{-10mm}= \frac{\sqrt{a}\,\delta_c(z)}{\sqrt{2\pi }S_n^{3/2}}{\rm e}^{-B_n^2/(2S)}\left[1+0.067\left(\frac{S_n}{a \delta_c^2(z)}\right)^{0.6}\right].\nonumber\\
&&
\label{FST}
\end{eqnarray}
As it is well-known (\cite{ST2002}), this first-crossing rate is not normalized to unity. This is a basic difference between the moving barrier  and the constant (spherical) barrier model. When the barrier height is constant, all random walks are guaranteed to cross the barrier because the
rms height of random walks at $S_n$ is proportional to $\sqrt{S_n}$. At sufficiently large $S_n$, all walks
will have crossed the constant barrier. In the moving barrier case, in which the barrier diverges when $S_n\rightarrow\infty$, not all trajectories intersect it. This is because the rms height of the random walk grows more slowly than the rate at which the barrier height increases and there is no guarantee that all random walks will intercept the barrier.
It seems reasonable to associate the fraction of random walks 
that do not cross the barrier with the particles that in N-body simulations are not associated to bound states (\cite{ST2002}).

After this rather long and technical summary of how to compute the
first-crossing rate for a generic moving barrier, we are ready to compute conditional probabilities.

\section{Halo Bias}
\label{sect:bias}

We now apply the technique of the previous section  to the computation of the halo bias,
including the non-Markovian corrections coming from the NG. We will use
a top-hat window function in wavenumber space. The calculation of the
non-Markovian effects on the bias from a top-hat window function  in real space can be found
in \cite{ma}.

\subsection{Conditional probability: the moving barrier case  and Gaussian initial conditions}

We begin our analysis with the simpler case in which the density field is Gaussian. Since we are also taking a top-hat filter in wavenumber space, the evolution is the Markovian.
To compute the bias, we need the probability of forming a halo of mass $M$,
corresponding to a smoothing radius $R$, under the condition that the
smoothed density contrast on a much larger scale $R_m$ has a specified
value $\d_m=\delta(R_m)$.  We use ${\cal F}_{\rm mb}(S_n|\d_m,S_m)$ to denote the
corresponding conditional first-crossing rate, i.e.,  the rate at which trajectories
first cross the barrier $\d=B(S)$ at time $S_n$, under the condition
that they passed through the point $\d=\d_m$ at an earlier time $S_m$.  We
also use the notation ${\cal F}_{\rm mb}(S_n|0)\equiv {\cal F}_{\rm mb}(S_n|\d_m=0,S_m=0)$,
so ${\cal F}_{\rm mb}(S_n|0)$ is the first-crossing rate when the
density approaches the cosmic mean value on very large scales.

The halo overdensity in Lagrangian space is given by
(\cite{MW96}; see also \citealt{Zentner} for a
review) 
\be 
   1+\d_{\rm halo}^L=\frac{{\cal F}_{\rm mb}(S_n|\d_m,S_m)}{{\cal
    F}_{\rm mb}(S_n|0)}\, .  
\ee 
The relevant quantity for our purposes is 
the halo conditional probability
 \begin{eqnarray} 
&&\hspace*{-6mm}\Pi_{\rm halo}(\delta_n,S_n| \delta_m,S_m)\equiv\label{Pbias}\nonumber\\
\hspace{-10mm}
&&\hspace*{-6mm}\frac{
\int_{-\infty}^{B_1} d\d_1\ldots \widehat{d\d}_m\ldots \int_{-\infty}^{B_{n-1}}d\d_{n-1}
W\left(\delta_0=0;\delta_1,\dots,\delta_n;S_n\right)}
{\int_{-\infty}^{B_1} d\d_1\ldots \int_{-\infty}^{B_{m-1}}d\d_{m-1}
W\left(\delta_0=0;\delta_1,\cdots,\delta_m;S_m\right)} ,\nonumber\\
&&
\label{Pihalogeneral}
\end{eqnarray}
where the   hat over $d\delta_{m}$ means that $d\delta_{m}$ must be omitted
from the list of integration variables.
The numerator is a sum over all
trajectories that start from $\d_0=0$ at $S=0$, have a given fixed value
$\d_m$ at $S_m$, and a value $\d_n$ at $S_n$, while all other points of the
trajectory, $\d_1, \ldots ,\d_{m-1},\d_{m+1},\ldots \d_{n-1}$ are
integrated up to the corresponding value of barrier, and we use the notation
$B_i\equiv B(S_i)$.  The denominator gives the
appropriate normalization to the conditional probability.

The conditional first-crossing rate
${\cal F}_{\rm mb}(S_n|\d_m,S_m)$ is obtained from the conditional probability
$\Pi_{\rm halo} (\delta_n,S_n| \d_m,S_m) $ using
\be\label{firstcrossTcond}
{\cal F}_{\rm mb}(S_n|\d_m,S_m)  = -\frac{\partial}{\partial S_n}\int_{-\infty}^{B_n}d\d_n\, 
 \Pi_{\rm halo} (\delta_n,S_n| \d_m,S_m)\, .
\ee
Since we are considering the Gaussian case, with a top-hat filter in wavenumber space, the probability density $W$ factorizes,
\bees\label{facto1}
&&W^{\rm gm}(\delta_0;\delta_1,\ldots, \delta_m, 
\ldots ,\delta_n; S_n)\nn\\
&=& 
W^{\rm gm}(\delta_0;\delta_1,\ldots, \delta_{m-1}, \delta_m;S_m)\nn\\
&&\times W^{\rm gm}(\delta_m; \delta_{m+1}, \ldots ,\delta_n; S_n-S_m)\, ,
\ees
and the halo probability $\Pi_{\rm halo} (\delta_n,S_n| \d_m,S_m) $ in \eq{Pbias} 
becomes identical to the
probability of arriving in $\d_n$ at time $S_n$, starting from $\d_m$ at
time $S_m$ for the moving barrier, reflecting the fact that the evolution of $\d(S)$ is in this case Markovian.
We can then compute $\Pi_{\rm halo}$
as in  the previous section, performing a  shift of the remaining integration variables 
$\d_i$ with
 $i=(m+1,\cdots,n-1)$,
\be\label{shift0}
\delta_i\rightarrow \delta_i-\sum_{p=1}^\infty\frac{B_n^{(p)}}{p!}
\, \left(S_i-S_n\right)^p\, ,
\ee
and we get 
\begin{eqnarray}
\label{pihalo}
&&\Pi_{\rm halo} (\delta_n,S_n| \d_m,S_m) =\nonumber\\
&&\frac{1}{\sqrt{2\pi (S_n-S_m)}}\,
\left(  {\rm e}^{-(\d_n-\d_m)^2/(2(S_n-S_m))}\right.\nonumber\\
&-& \left.{\rm e}^{-(2B_n-\d_n-\d_m)^2/(2(S_n-S_m))} \right)\nonumber\\
&+&{2(B_n-\delta_n)\over \sqrt{2\pi} (S_n-S_m)^{3/2}}
 {\rm e}^{-(2B_n-\delta_n-\d_m)^2/(2 (S_n-S_m))}{\cal P}_{mn}\nn\\
&-&{2(B_n-\delta_n)^2\over \sqrt{2\pi} (S_n-S_m)^{5/2}}  
 {\rm e}^{-(2B_n-\delta_n-\d_m)^2/(2 (S_n-S_m))}
{\cal P}^2_{mn}\,,\nonumber\\
&&
\end{eqnarray}
where
\be
{\cal P}_{mn}\equiv{\cal P}(S_m,S_n)=\sum_{p=1}^\infty\frac{(S_m-S_n)^p}{p!}B_n^{(p)}\, .
\ee
In the following, we will use this quantity with the  sum truncated to $p=5$, as discussed in Section \ref{sect:pi} and as advocated by \cite{ST2002} for the 
conditional mass function.

The calculation of the conditional first-crossing rate ${\cal F}_{\rm mb}(S_n|\d_m,S_m)$ proceeds by  
taking the derivative with respect to $S_n$ 
\be
{\cal F}_{\rm mb}(S_n|\d_m,S_m)=
\frac{(B_n-\d_m)+\mathcal{P}(S_m,S_n)} {\sqrt{2\pi}(S_n-S_m)^{3/2}} {\rm e}^{-{(B_n-\d_m)^2\over 2(S_n-S_m)}}\, .
\ee
In a sufficiently large region $S_m\ll S_n$ and $\d_m\ll \d_n$.
Then, expanding to quadratic order in   $\d_m$ and
after mapping to Eulerian space, we find the first two
Eulerian bias coefficients

\be
\label{d1haloG}
b_1 \simeq
  1+\frac{B_n}{S_n}-\frac{1}{B_n+{\cal P}(0,S_n)}\, .
  \ee
and
\be
\label{d2haloG}
b_2 \simeq \frac{B^2_n}{S_n^2}-\frac{1}{S_n}-2
\frac{B_n}{(B_n+{\cal P}(0,S_n))S_n}\, .
\ee
The above results hold for a generic barrier. We now examine it for different collapse models.

\subsubsection{Constant barrier}

We first apply these results to  the spherical collapse model where the barrier is  constant $\delta_c(z)$. 
Using the standard  notation $\nu\equiv \delta_c(z)/\sigma$  we  get, for the bias coefficients, 
\be
b_1
   \simeq
  1+\frac{\nu^2}{\delta_c}- \frac{1}{\delta_c}\, ,
  \ee
 and
\be
b_2\simeq
 \frac{\nu^4}{\delta^2_c}
 - 3\frac{\nu^2}{\delta_c^2}\, .
\ee
\subsubsection{Ellipsoidal barrier}

For the   the ellipsoidal
barrier (\ref{ST}),  we  get, for the bias coefficients, 
\begin{eqnarray}
 b_1 &\simeq&
  1+\sqrt{a}\frac{\nu^2}{\delta_c}\left[1+0.4\left(\frac{1}{a\nu^2}\right)^{0.6}\right]
 \nonumber\\
 &-& \frac{1}{\sqrt{a}\delta_c\left[1+0.067\left(\frac{1}{a\nu^2}\right)^{0.6}\right]}\, .
 \label{deltahalo1}
  \ees
 and
\begin{eqnarray}
b_2&\simeq&
 a\frac{\nu^4}{\delta^2_c}\left[1+0.4\left(\frac{1}{a\nu^2}\right)^{0.6}\right]^2
 - \frac{\nu^2}{\delta_c^2}\nonumber\\
 &-&2\frac{\nu^2}{\delta_c^2}\frac{1+0.4\left(\frac{1}{a\nu}\right)^{0.6}}{1+0.067\left(\frac{1}{a\nu^2}\right)^{0.6}}\,.
\ees
For large masses, $\nu^2\gg 1$, the bias coefficient scales like
\be
 b_1\simeq \sqrt{a}\,\frac{\nu^2}{\delta_c}\, .
 \label{kk}
 \ee
Observe that this result differs from that proposed by  \cite{SMT}, which in the large mass limit
rather  scales like $b_1\sim a(\nu^2/\delta_c)$, i.e. it is smaller than
(\ref{kk}) by a  factor $\sqrt{a}$. The proportionality to $\sqrt{a}$ in \eq{kk} can be traced to the fact that, in the large mass limit, $b_1$ is dominated by the term $B_n/S_n$ in \eq{d1haloG}
and, for the barrier given in eq.~(\ref{ST}), $B_n$ is proportional to $\sqrt{a}$. In other words, this dependence is a consequence of the fact that, in the large mass limit, the barrier of  \cite{SMT} does not reduce to the spherical collapse model barrier $\d_c$, but rather to $\sqrt{a}\d_c$, so the large mass limit (\ref{kk}) can be formally obtained from the spherical collapse model by performing the replacement 
$\d_c\ra\sqrt{a}\d_c$. For \cite{SMT}, the physics of spherical collapse always gives $\delta_c$, 
not $\sqrt{a} \delta_c$, so it 
is really only the variances 
which were rescaled, $\sigma^2 \rightarrow \sigma^2/a$.  Since the unconditional mass functions are the same in our case and in the one discussed by \cite{SMT}, 
but the bias factors are not, this means that one
could, in principle, discriminate between whether $a$
should be thought of as rescaling $\delta_c$  or $\sigma^2$. 
 
\begin{figure}
\centering
\includegraphics[width=0.4\textwidth]{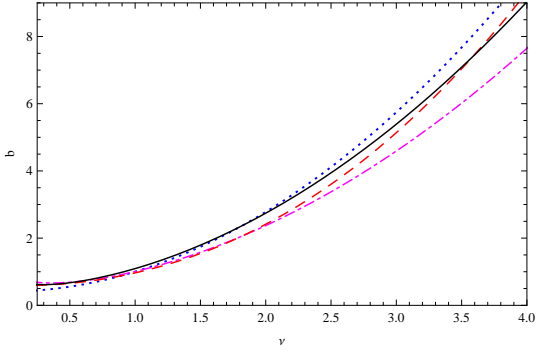}
\caption{
\label{fig:bias}
The halo bias at linear order as a function of $\nu=\delta_c/\sigma$.
Our formula in Eq.~(\ref{deltahalo1})  (solid black line) is compared to other
results: the fit  to N-body simulations as in  \protect\cite{Tinker2010}  (dashed red), and 
two standard approximations: spherical (dotted blue) and  the SMT expression for the bias (dot-dashed magenta),
which is also predicted by the diffusing barrier model of Section \ref{subsec:diffbarrier}.
}
\end{figure}

In Fig.~\ref{fig:bias} we show our result (\ref{deltahalo1}) for the bias and compare it to the fit
to N-body simulations of \cite{Tinker2010} and to 
two standard approximations: spherical collapse and the SMT result.
Our result for the bias is manifestly in better agreement with N-body simulations than the 
commonly-used approximations. 

\subsubsection{The diffusing barrier}  
\label{subsec:diffbarrier}

In MR2 it has been proposed a model in which the barrier performs a diffusing motion, with diffusion coefficient $D_B$, around an average value which for the spherical collapse model is simply $\d_c(z)$ (without any factor of $\sqrt{a}$) while for the ellipsoidal model is given 
\be\label{STnoa}
B(S_n)\simeq \delta_c(z)\[ 1 +0.4\(\frac{S_n}{a\delta_c^2(z)}\)^{0.6}\]\, ,
\ee
i.e. by \eq{ST} without the factor $\sqrt{a}$. \Eq{STnoa} is in fact the barrier which actually follows for an ellipsoidal model for collapse, while the factor $\sqrt{a}$ in \eq{ST} was simply inserted by hand, in order to fit the data. Including the stochastic motion of the barrier, which is meant to mimic a number of random effects in the process of halo formation, it was found in MR2 that the relative motion of the trajectory $\d(S)$ and of the barrier is a stochastic motion with diffusion constant $(1+D_B)$ so that, for instance, in the Markovian case the evolution of the probability distribution is governed by a Fokker-Planck equation of the form
\be\label{FPDB}
\frac{\partial \Pi_{\rm mb} (\d_n;S_n)}{\partial S_n}=\frac{(1+D_B)}{2}\frac{\partial^2\Pi_{\rm mb} (\d_n;S_n)}{\partial\delta^2}\, .
\ee
The result for this diffusive barrier can therefore be obtained from the result for the spherical collapse barrier, or from the result for the barrier (\ref{STnoa}), by formally rescaling
$S\ra (1/a)S$, where $a=1/(1+D_B)$. 
For the halo mass function, which depends only on the combination $\nu= \d_c/\s=\d_c/\sqrt{S}$, the rescaling $S_n\ra (1/a)S_n$ is equivalent to the rescaling $\d_c\ra \sqrt{a}\d_c$, and therefore the diffusive barrier model produces the same result as that obtained with the SMT barrier (\ref{ST}). 

This equivalence, however, does not extend to the halo bias. In fact, in the large mass limit, the halo bias deduced using the barrier (\ref{STnoa}) in \eq{d1haloG} is
$b_1\simeq (\nu^2/\delta_c)=\d_c/S_n$, and rescaling $S_n\ra (1/a)S_n$ gives
\be
 b_1\simeq a\, \,\frac{\nu^2}{\delta_c}\, ,
 \label{kk2}
 \ee
which differs from \eq{kk} by a factor $\sqrt{a}$ and agrees with \cite{SMT}.

\subsection{Adding non-Markovianity from a top-hat window function in real space,
in the high mass limit}
\noindent
Furthermore, one should add to this result the non-Markovian corrections due to the use of a top-hat filter function in coordinate space. In the large mass limit, where
\eq{STnoa} reduces the  constant barrier, the computation of the bias was performed in \cite{ma},
where it was found
that $ b_1$ gets multiplied by a factor $1/(1-a\kappa)$, so for the diffusive barrier model with Markovian corrections due to the filter we get, in the large mass limit,
\be\label{kkdiff}
 b_1\simeq \frac{a}{1-a\kappa}\,\frac{\nu^2}{\delta_c} 
 =\(\frac{\sqrt{a}}{1-a\kappa}\)\,\sqrt{a}\,\frac{\nu^2}{\delta_c} 
 \, ,
\ee
where $\kappa$ is a parameter that controls the non-Markovian effects of the filter function, whose numerical value can be estimated as in MR1, but which for accurate fitting  is better treated as a free parameter, see \cite{ma}.
Observe that in \cite{ma} the spherical collapse result was rescaled according to $\d_c\ra \sqrt{a}\d_c$, which actually corresponds to the replacement in the SMT barrier model and not to the diffusive barrier model, so the resulting value for the bias was  incorrectly obtained to be $b_1\simeq [\sqrt{a}/(1-a\kappa)]
(\nu^2/\delta_c)$. 

In particular, consider that the rescaling
$\kappa\ra a \kappa$, which leads to the factor $(1-a \kappa)$ in
the denominator of eq.~(\ref{kkdiff}), is obtained assuming that the
barrier performs a simple diffusive Markovian motion. In a
realistic description, the actual stochastic motion can be more
complicated and the reader should be aware of the limitations of this procedure. Unfortunately,
we  have not been able to find an analogue of the expression (\ref{kkdiff})
for small masses.

\subsection{Conditional probability: the moving barrier case and non-Gaussian initial conditions}

Deviations from Gaussianity are  encoded, e.g., in the connected 
 three- and four-point correlation functions which are dubbed the bispectrum
and the trispectrum,
respectively. A phenomenological way of parametrizing the level of NG is to
expand the fully 
 non-linear primordial Bardeen
 gravitational potential $\Phi$ in powers of the linear gravitational potential
$\Phi_{\rm L}$
 \be
 \label{phi}
 \Phi=\Phi_{\rm L}+f_{\rm NL}\left(\Phi_{\rm L}^2-\langle\Phi_{\rm
L}^2\rangle\right)\, .
 \ee
The  dimensionless
quantity  $f_{\rm NL}$  sets the
magnitude of the three-point 
correlation function  (\cite{bartoloreview}). If the process generating the primordial
NG  
is local in space, the parameter  $f_{\rm NL}$ in Fourier
space is
independent of the momenta
entering the corresponding  correlation functions; if instead the process which
generates the
primordial cosmological perturbations is
non-local in space, like in 
models of inflation with non-canonical kinetic terms, $f_{\rm NL}$ 
acquires a dependence on the momenta. 
The strongest current
limits on the strength of local NG set the $f_{\rm NL}$
parameter to be in the range $-4<f_{\rm NL}<80$ at 95\% confidence level
\citep{zal}. 
The goal of this subsection is to compute the halo bias parameters  in the presence of NG and for the ellipsoidal collapse, using the technique developed in MR3, which generalizes
excursion set theory to deal with non-Gaussianities in the primordial density field. We will also
work with a top-hat window function in real space.

Similarly to the Gaussian  case, the probability of arriving in $\d_n$ in a
``time'' $S_n$, starting from the initial value $\delta_0=0$, without ever
going above the threshold, in the presence of NG is given by 
\begin{eqnarray}
\Pi_{\rm mb} (\d_n;S_n)
& \equiv&\int_{-\infty}^{B_1} d\d_1\ldots \int_{-\infty}^{B_{n-1}}d\d_{n-1}\, \nn\\
&&\times W_{\rm NG}(\d_0;\d_1,\ldots ,\d_{n-1},\d_n;S_n).
\label{NG}
\end{eqnarray}
where 
\begin{eqnarray}
\label{aa}
&&W_{\rm NG} (\d_0;\d_1,\ldots ,\d_n;S_n)=\int{\cal D}\lambda\nn\\
&&\times\exp\left\{ i\sum_{i=1}^n\lambda_i\d_i-\frac{1}{2}\sum_{i,j=1}^n
\lambda_{i}\lambda_{j}\,
{\rm min}(S_i,S_j)\right\}\nn\\
&&\times\exp\left\{\frac{(-i)^3}{6}\sum_{i,j,k=1}^{n}\langle\d_i\d_j\d_k\rangle_c\lambda_i\lambda_j\lambda_k
\right\}\, ,
\end{eqnarray}
and we have retained only the three-point connected correlator $\langle\delta_i\delta_j\delta_k\rangle_c$ as a signal of NG. It is now clear where the non-Markovianity is coming from when
non-Gaussian initial conditions are present: expanding $W_{\rm NG}$ in powers of
$\sum_{i,j,k=1}^{n}\langle\d_i\d_j\d_k\rangle_c\lambda_i\lambda_j\lambda_k$ and going to the continuum, one obtains integrals over the intermediate times which
introduce memory effects.

As in eq.~(\ref{Pihalogeneral}), the conditional probability relevant to the
halo bias is obtained keeping $\delta_m$ fixed, rather than
treating it as an integration variables. 
In principle the non-Markovian contribution to the halo probability from NG   should be  computed separating the various contributions to the sum
according to whether an index is  smaller than or equal to $m$, larger than $m$ and smaller than or equal to $n$. Fortunately, to compute the halo bias parameters, at the end one needs to take
the limit $S_m\ll S_n$ and $\delta_m\ll \d_n$. 
In this limit we can the safely neglect the contribution to the sum from
all indices running from $1$ to $m$, since in general $\langle\d_m^p\d_n^q\rangle$ scales like
$(S_m^{p/2}S_n^{q/2})$, where $p,q\geq 0$ and $p+q=3$, and therefore vanishes for $S_m\to 0$. 
In other words, in \eq{aa} we can replace 
$\sum_{i,j,k=1}^{n}$ with $\sum_{i,j,k=m+1}^{n}$.  As usual, we then expand the exponential and use the fact that $i\lambda_k\exp\{i\sum_i\lambda_i\delta_i\}=\pa_k\exp\{i\sum_i\lambda_i\delta_i\}$, and we use the factorization property (\ref{facto}) (\cite{MR3}). This gives
\bees
&&W_{\rm NG} (\d_0;\d_1,\ldots ,\d_n;S_n)\nn\\
&\simeq&
\[ 1-\frac{1}{6}\sum_{i,j,k=m+1}^{n}\langle\d_i\d_j\d_k\rangle_c\pa_i\pa_j\pa_k\]\nn\\
&&W^{\rm gm}(\delta_0;\delta_1,\ldots, \d_m;S_m)\nn\\
&&\times W^{\rm gm}(\d_m; \delta_{m+1}, \ldots ,\delta_n; S_n-S_m)\, .
 \ees
Since  now the derivatives act only on the second $W$ factor, the first $W$ factor factorizes and, after integration over $\d_1,\ldots, \d_{m-1}$,  cancels the denominator in the conditional probability
(\ref{Pbias}).

Next, as in MR3, we  introduce the notation
\begin{eqnarray}
&G_3^{(p,q,r)}(S_n)\equiv&\nn\\
&\[\frac{d^p}{dS_i^p}\frac{d^q}{dS_j^q}\frac{d^r}{dS_k^r}
\langle\d(S_i)\d(S_j)\d(S_k)\rangle_c\]_{S_i=S_j=S_k=S_n}.&
\end{eqnarray}
and we expand the correlator as 
\bees
\label{pqrs}
\langle\d(S_i)\d(S_j)\d(S_k)\rangle
&=&
\sum_{p,q,r=0}^{\infty} \frac{(-1)^{p+q+r}}{p!q!r!}(S_n-S_i)^{p}\\
&\times&
(S_n-S_j)^{q}(S_n-S_k)^{r}G_3^{(p,q,r)}(S_n)\, .\nn
\ees
As shown in MR3, in the large mass limit (which is the most interesting regime for observing the non-Gaussianities) the leading contribution to the halo bias probability
is given by the term in \eq{pqrs} with $p=q=r=0$. We neglect subleading contributions, which can be computed with the same technique developed in MR3. The discrete sum then reduces to
$\langle\delta_n^3\rangle_c \sum_{i,j,k=m+1}^{n}\partial_i\partial_j\partial_k$. 
We can now peoceed as in the previous section and  
perform the  shift of variables (\ref{shift}) on $\d_i$, for with $i=m+1,\cdots,n-1$. The halo probability becomes
\begin{eqnarray}
&&\Pi_{\rm halo\, NG} (\delta_n,S_n| \d_m,S_m) =\nn\\
&&\int_{-\infty}^{B_n} d\d_{m+1}\ldots \int_{-\infty}^{B_n}d\d_{n-1}\, \nn\\
&&{\rm e}^{-\frac{1}{6}\langle\d_n^3\rangle_c \sum_{i,j,k=m+1}^{n}\partial_i\partial_j\partial_k}
W_{\rm mb}(\d_m;,\ldots ,\d_{n-1},\d_n;S_n)\, ,\nonumber\\
&&
\end{eqnarray}
where  $W_{\rm mb}$ is the probability density in the space of trajectories  with a moving barrier, so that,
in general,
\begin{eqnarray}
\label{aaaaa}
&&\hspace*{-5mm}\int_{-\infty}^{B_n} d\d_{m+1}\ldots \int_{-\infty}^{B_n}d\d_{n-1}\, 
W_{\rm mb}(\d_m;\d_{m+1},\ldots ,\d_{n-1},\d_n;S_n)\nn\\
&&\hspace*{-5mm}=\Pi_{\rm mb}(\d_m;\d_n; S_n)=
\Pi^{(0)}_{\rm mb} +\Pi^{(1)}_{\rm mb} +\Pi^{(2)}_{\rm mb} +\cdots\, ,
\end{eqnarray}
where, as in \eq{Pi0Pi1Pi2}, the terms $\Pi^{(0)}_{\rm mb}$, $\Pi^{(1)}_{\rm mb}$, 
$\Pi^{(2)}_{\rm mb}$, etc. correspond to the different orders in the expansion of the exponential
in \eq{eq22next}, but now for trajectories that start at $\d_m$, with $\d_m$ small but finite,
 rather than at $\d_0=0$.

To compute these expressions we can now use the following identity, proven in MR3,
\begin{eqnarray}
&&\sum_{i,j,k=1}^{n}
\int_{-\infty}^{B_n} d\d_1\dots d\d_{n}\,
\pa_i\pa_j\pa_k W_{\rm mb}\nn\\
&&=\frac{\pa^3}{\pa B_n^3}\int_{-\infty}^{B_n}d\d_n\, \Pi_{\rm halo} (\delta_n,S_n| \d_m,S_m)\, ,
\end{eqnarray}
where $\Pi_{\rm halo} (\delta_n,S_n| \d_m,S_m)$ is given in eq.~(\ref{pihalo}).
The calculation of the conditional first-crossing rate ${\cal F}_{\rm mb\, NG}(S_n|\d_m,S_m)$ proceeds by finally 
taking the derivative with respect to $S_n$:
\begin{eqnarray}
&&\hspace*{-5mm}
{\cal F}_{\rm mb\,NG}(S_n|\d_m,S_m)=
\frac{(B_n-\d_m+{\cal P}_{mn})}{\sqrt{2\pi}(S_n-S_m)^{3/2}}
 {\rm e}^{-{(B_n-\d_m)^2\over 2(S_n-S_m)}}
\nn\\
&&\hspace*{-5mm}+\frac{{\cal S}_3}{6\sqrt{2\pi }(S_n-S_m)^{5/2}}
\left[(B_n-\delta_m)^4
\right.\nn\\
&& \hspace*{-5mm}
-(B_n-\delta_m)^3({\cal P}_{mn}+2(S_n-S_m) B_n')
\nn\\
&&\hspace*{-5mm} +2(B_n-\delta_m)^2
\left(-(S_n-S_m)+{\cal P}_{mn}^2+(S_n-S_m){\cal P}_{mn}B_n' \right) \nn\\
&&\hspace*{-5mm} +(S_n-S_m) (B_n-\delta_m)({\cal P}_{mn}+6(S_n-S_m) B_n' \nn\\
&& \hspace*{-5mm} -4{\cal P}_{mn}^2B_n'-2(S_n-S_m){\cal P}_{mn}')-(S_n-S_m)^2 \nn\\
&& \hspace*{-5mm}  -2(S_n-S_m){\cal P}_{mn}({\cal P}_{mn}+(S_n-S_m) B_n'
\nonumber\\
&&\hspace*{-5mm}-\left. 4(S_n-S_m){\cal P}_{mn}')\right]{\rm e}^{-{(B_n-\d_m)^2\over 2(S_n-S_m)}}\nn\\
&&\hspace*{-5mm}+\frac{(S_n-S_m)^2{\cal S}'_3}{3\sqrt{2\pi }(S_n-S_m)^{5/2}}\left[ (B_n-\delta_m)^2-(B_n-\delta_m){\cal P}_{mn}
\right.\nn\\
&&\hspace*{-5mm}-\left.(S_n-S_m)+2{\cal P}_{mn}^2\right] {\rm e}^{-{(B_n-\d_m)^2\over 2(S_n-S_m)}}\,,
\label{fmbng}
\end{eqnarray}
where the prime stands for derivative with respect to $S_n$. Notice that in the limit of vanishing
$f_{\rm NL}$ only the first line of the above equation survives.
It should also be observed that terms proportional to derivatives of the cumulant  such as ${\cal S}'_3$ (which are subleading compared to the terms proportional to ${\cal S}_3$)
can also come from terms with $p+q+r\geq 1$ in \eq{pqrs}, that we have neglected, buth which in principle can be computed as in MR3.
Expanding again in powers of $\delta_m$ up to $\d_m^2$, we find the scale-independent NG contribution to the Eulerian 
bias parameters. Normalizing  the bispectrum as
\be\label{defbis}
{\cal S}_3(S_n)\equiv \frac{1}{S_n^2}\langle\d^3(S_n)\rangle\, ,
\ee
we find the corrections to the halo bias parameters from NG
\begin{eqnarray}
&&\hspace*{-8mm} \Delta b_{1 \rm\,NG} \simeq 
   -\frac{{\cal S}_3(S_n)}{6 S_n(B_n+{\cal P}_n)^2}\left[
  3B_n^4+2B_n^3({\cal P}_n-2S_nB'_n)\right.\nonumber\\
  &-&B_n^2(2S_n+{\cal P}_n^2+4 S_n{\cal P}_n B_n')\nonumber\\
  &+&4 B_n{\cal P}_n(-S_n+{\cal P}_n^2+S_n{\cal P}_nB'_n)\nonumber\\
  &+&\left.S_n(S_n+{\cal P}_n(3{\cal P}_n+8 S_nB'_n-4{\cal P}_n^2 B'_n
  -10S_n{\cal P}'_n)\right]\nn\\
  &-&{{\cal S}'_3(S_n) S_n \over 3(B_n+{\cal P}_n)^2}
  \left(
  (B_n+3{\cal P}_n)(B_n-{\cal P}_n)+S_n
  \right)\, ,
   \label{d1haloNG}
 \end{eqnarray}
and
\begin{eqnarray}
&&\hspace*{-8mm} \Delta b_{2 \rm\,NG} \simeq 
   \frac{{\cal S}_3(S_n)}{3 S_n^2 (B_n+{\cal P}_n)^2}\left[
  -3B_n^5+B_n^4(-2{\cal P}_n+4 B'_n S_n)\right.\nonumber\\
  &+&B_n^3({\cal P}^2_n+4 S_n{\cal P}_nB_n'+8 S_n)\nonumber\\
   &+&\hspace*{-3mm}B_n^2(7{\cal P}_n S_n-4{\cal P}_n^3-6 S_n^2 B'_n-4 S_n{\cal P}_n^2 B_n')\nonumber\\
  &+&\hspace*{-3mm} B_n (4 S_n {\cal P}_n^3 B'_n -12 S_n^2 {\cal P}_n B'_n+10 S_n^2 {\cal P}_n {\cal P}'_n
     \nonumber\\
  &-&\hspace*{-3mm}\left. 
  S_n(3S_n+4{\cal P}_n^2))+2 {\cal P}_n S_n({\cal P}_n^2-S_n)+2 {\cal P}_n^2 S_n^2 B'_n \right]\nn\\
  &-&\hspace*{-3mm}{2{\cal S}'_3(S_n)  \over 3(B_n+{\cal P}_n)^2}
  \left(
  B_n(B_n+3{\cal P}_n)(B_n-{\cal P}_n)-S_n{\cal P}_n
  \right)\, , 
 \label{d2haloNG} 
 \end{eqnarray}
where ${\cal P}_n={\cal P}(0,S_n)$.  
As shown in Fig.~\ref{fig:DeltabNG}, these terms correct the bias calculated with Gaussian initial conditions
by a few percents.
In the last term we have expanded for large masses and again assumed that the barrier as well as ${\cal S}_3$ are  slowly varying.
At first one might think that such an approximation, altough useful in some cases, would not apply to  the barrier which corresponds to the the ellipsoidal collapse, \eq{ST}. In this case in fact $B_{\rm ST}(S_n)$ is given by a constant plus a term proportional to $S_n^{\g}$ with $\g\simeq 0.6<1$, and therefore already its first derivative, which is proportional to $S_n^{\g-1}$ is large at sufficiently small $S_n$, and formally even diverges as $S_n\ra 0$. However one should not forget that, in practice, even the largest galaxy clusters than one finds in observations, as well as in large-scale $N$-body simulations, have typical masses smaller than about $10^{15} h^{-1}\msun$ which, in the standard $\Lambda$CDM
cosmology, corresponds to values of $S_n\gsim 0.35$, see e.g. Fig.~1 of \cite{Zentner}.  Even for such a value, which is the smallest  we are interested in, the value of $B'_{\rm ST}(S_n)$ is just of order 0.3 which means that, in the range of masses of interest, the barrier of ellipsoidal collapse can be considered as slowly varying. Note that  we have assumed a barrier shape which is the same for Gaussian initial conditions. A justification of this assumption can be found in \cite{lamd}.

\begin{figure}
\centering
\includegraphics[width=0.4\textwidth]{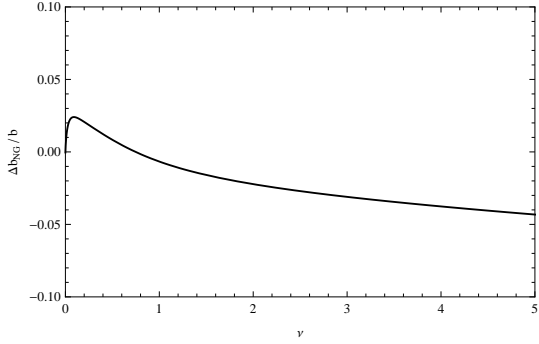}
\caption{
\label{fig:DeltabNG}
The magnitude of the NG correction (with $f_{\rm NL}=100$) to the first bias parameter (Eq.~(\ref{d1haloNG})), normalized to the halo bias without NG (Eq.~(\ref{d1haloG})), 
 as a function of $\nu=\delta_c/\sigma$.
}
\end{figure}

For high mass halos, $\nu^2\gg 1$,  the scale-independent NG contribution to the Eulerian 
bias parameters are
\begin{eqnarray}
\Delta b_{1\rm\,NG} &\simeq& 
   -\frac{1}{6}{\cal S}_3\left[3a\,\nu^2-1.6\,(a\nu^2)^{0.4}\right]\,,\nonumber\\
   \Delta b_{2\rm \,NG} &\simeq& 
   -\frac{1}{3}{\cal S}_3\left[3a^{3/2}\frac{\nu^4}{\delta_c}-1.6\,a^{0.9}\frac{\nu^{2.8}}{\delta_c}\right]\,.
   \ees
Notice that the leading term of the first bias coefficient does not agree with what found by \cite{marian}, 
who found $\Delta b_{1 \rm\,NG} \simeq
   -(2/6){\cal S}_3\nu^2$ for high mass halos, even in the limit in which we reduce artificially ourselves to the spherical
collapse case ($a=1$) adopted  to produce their formula (39). It might well be  that the discrepancy arises from the fact that \cite{marian} deduce  the halo bias coefficient from  the NG halo
mass function defined artificially as the Gaussian Sheth-Tormen mass function multiplied by the ratio of the Press-Schecter  NG and the Gaussian mass functions. On the other hand, had we also considered
the variance $S_n$ and the skewness ${\cal S}_3$ as a function of scale through the long wavelength
model $\delta_m({\bf k})$, we could have obtained as well the standard scale-dependent part of the
NG bias coefficient, see \cite{fabian}.

\section{The two-barrier first-crossing rate}
\label{sect:2barrier}

Following \cite{lc}, we now wish to calculate through the excursion set method the
two-barrier first-crossing rate.
Before launching ourselves into the computation of the conditional probability, we pause to make 
some preliminary  considerations. In particular, Section 4.2 will be devoted to the case of Gaussian
initial conditions and our aim is there to provide an analytical understanding of the problem previously discussed by \cite{Bond};  the case of NG initial conditions will be discussed in Section 4.3. 
\cite{Bond}  showed that the uncorrelated
steps solution for the two-barrier problem appears to be in better
agreement with halo conditional mass
functions than did the solution for correlated steps. 
Given this, it is natural to ask if going from constant
to moving barriers alleviates the discrepancy.
This is excellent motivation for studying the two-moving
barriers problem.  In this case, the uncorrelated
steps problem was solved numerically by \cite{ST2002} who also provided an analytic
approximation, which will will discuss below. 
Moreover, \cite{ST2002} also showed that these walks (with uncorrelated steps) and their associated
formulae for the first crossing distribution,
were able to provide a rather good description 
of the conditional halo mass function. 
So the most interesting question would be if 
accounting for correlated steps helps in the computation of the two-barrier problem with moving barriers. 
We will not discuss this problem combining Gaussian initial conditions with non-Markovian evolution of the noise for moving barriers. 
However, we will include the non-Markovianity induced by the NG initial conditions.

\subsection{Some preliminaries}
\label{preliminiaries}

Let us first analyze the problem with only one barrier, $B_n$. An instructive
alternative way of understanding why the first crossing rate is obtained 
by taking (minus) the  derivative with respect to $S_n$ is the following. In order
to impose that
the trajectory makes its first barrier crossing at ``time" $S_n$,
we require that for all times $S_1,\ldots S_{n-1}$ the trajectory  stays
below $B_n$, while at $S_n$ it must be above. Therefore the quantity
that we need is
\be\label{eq1}
\int_{-\infty}^{B_1}d\d_1\ldots \int_{-\infty}^{B_{n-1}}
d\d_{n-1}
\int_{B_n}^{\infty}d\d_n\,
W(\d_0;\d_1,\ldots,\d_n;S_n)\, .
\ee
We now write
\be\label{eq2}
\int_{B_n}^{\infty}d\d_n\,
=\int_{-\infty}^{\infty}d\d_n\,
-\int_{-\infty}^{B_n}d\d_n
\ee
and  use the fact that 
\bees
&&\hspace*{-5mm}\int_{-\infty}^{B_1}d\d_1\ldots \int_{-\infty}^{B_{n-1}}d\d_{n-1}
\int_{-\infty}^{\infty}d\d_n\,
W(\d_0;\d_1,\ldots,\d_n;S_n)\nn\\
&=&\int_{-\infty}^{B_n}d\d_1\ldots \int_{-\infty}^{B_{n-1}}d\d_{n-1}
W(\d_0;\d_1,\ldots,\d_{n-1};S_{n-1})\nn\\
&=&\int_{-\infty}^{B_{n-1}}d\d_{n-1}
\Pi_{\rm mb}(\d_0;\d_{n-1};S_{n-1})\, .
\ees
The second contribution obtained from \eq{eq2}
is
\bees
&&\int_{-\infty}^{B_1}d\d_1\ldots\int_{-\infty}^{B_{n}} d\d_{n}
W(\d_0;\d_1,\ldots,\d_n;S_n)\nn\\
&=&\int_{-\infty}^{B_n}d\d_{n}\Pi_{\rm mb}(\d_0;\d_n;S_{n})\, .
\ees
Then we get
\bees
&&\hspace*{-5mm}\int_{-\infty}^{B_1}d\d_1\ldots \int_{-\infty}^{B_{n-1}}d\d_{n-1}
\int_{B_n}^{\infty}d\d_n\,
W(\d_0;\d_1,\ldots,\d_n;S_n)\nn\\
&=&\hspace*{-3mm}\int_{-\infty}^{B_{n-1}}d\d_{n-1}
 \Pi_{\rm mb}(\d_0;\d_{n-1};S_{n-1})\nn\\
 &&-\int_{-\infty}^{B_n}d\d_n\Pi_{\rm mb}(\d_0;\d_n;S_{n})\nn\\
&=&\hspace*{-3mm}-\eps\frac{\partial}{\partial S_n}\int_{-\infty}^{B_n}d\d_n \Pi_{\rm mb}(\d_0;\d_n;S_{n})\nn\\
&&-\int_{B_{n-1}}^{B_n}d\d_{n-1}\, \Pi_{\rm mb}(\d_0;\d_{n-1};S_{n-1})
+{\cal O}(\eps^2)
\label{eq:der}
\, ,
\ees
where $\eps=(S_n-S_{n-1})$. The integral in the last line in the continuum limit becomes
$(B_n-B_{n-1})\Pi_{\rm mb}(\d_0;B_n;S_{n})$ and, since $(B_n-B_{n-1})=O(\eps)$ while
$\Pi_{\rm mb}(\d_0; \d_n=B_n;S_n)$ vanishes as $\sqrt{\eps}$, this term is overall ${\cal O}(\eps^{3/2})$, while the term proportional to $\pa/\pa S_n$ is ${\cal O}(\eps)$. Therefore the transition rate per unit time-step
$\eps$ is 
\be\label{eqF}
{\cal F}_{\rm mb}(S_n)=-\frac{\partial}{\partial S_n}\int_{-\infty}^{B_n}d\d_n\,
\Pi_{\rm mb}(\d_0;\d_n;S_{n})\, ,
\ee
which is the standard result.

In the two-barrier problem, denoting by $B^a(S_m)$ and $B^b(S_n)$ the 
two barriers corresponding to redshift $z_a$ and halo mass $M_m$ and $z_b$ and $M_n$, respectively,  the relevant quantity is the conditional probability given by
the ratio between

\bees\label{num}
{\cal  N}&\equiv&\int_{-\infty}^{B^a_1}d\d_1\ldots \int_{-\infty}^{B^a_{m-1}}d\d_{m-1}
\int_{B^a_m}^{B^b_m}d\d_m\nonumber\\
&&\times
\int_{-\infty}^{B^b_{m+1}}d\d_{m+1}\ldots \int_{-\infty}^{B^b_{n-1}}d\d_{n-1}\nonumber\\
&&\times
\int_{B^b_n}^{\infty}d\d_n\,
W(\d_0;\d_1,\ldots \d_n;S_n)\, ,
\ees
and
\bees
{\cal D}&\equiv&\int_{-\infty}^{B^a_1}d\d_1\ldots \int_{-\infty}^{B^a_{m-1}}
d\d_{m-1}\nn\\
&&\times \int_{B^a_m}^{B_m^b}d\d_m\,
W(\d_0;\d_1,\ldots \d_{m};S_{m})\, .\label{den}
\ees
In \eq{num} the integral over $d\d_m$ has a lower limit  $B^a_m$
because we require that the trajectory crosses above this barrier. The
subsequent evolution can bring it below this barrier again, so the
integrals over $d\d_{m+1}, \ldots d\d_{n-1}$ have as lower limit
$-\infty$.
Furthermore,  
the upper integration limit for the integrals over 
$d\d_{m}, \ldots ,d\d_{n-1}$ is given by the upper barrier $B^b$
because we want to sum only over  trajectories that  never crossed the
second barrier $\d_n$ at times smaller than $S_n$. 
Observe that this must be imposed even in the integral
over $d\d_m$.  Finally, at $S_n$, the trajectory crosses for the first
time above $B_n$, so the corresponding integral runs from $B^b_n$ to $+\infty$.
The denominator (\ref{den}) gives the appropriate normalization to the conditional probability, so that in the Markovian case, where $W$ factorizes, it cancels against the integrations over
$d\d_1,\ldots d\d_m$ in the numerator.
 Observe also that the
integral over $d\d_m$ (both in the numerator and in the denominator) in the continuum limit can equivalently written as in integral from $B_m^a$ and $+\infty$,
since at time $S_{m-1}$ the trajectory was below the lower barrier $B^a$, and the probability that in an infinitesimal time step $\eps$ it jumps above the upper $B^b$
vanishes as $\exp\{-(B^b_m-B^a_{m-1})^2/(2\eps)\}$, so it does not contribute to the continnum limit (at least, as long as there is a finite separation between the two barriers).

We now wish to derive a result analogous to \eq{eqF} for the conditional 
two-barrier first-crossing rate, i.e. for the rate at which trajectories first cross the upper barrier $B^b$ at $S=S_n$, under the condition that they first crossed the lower barrier $B^a$ at $S=S_m$.
In the Markovian case we can repeat  the derivation of \eqst{eq1}{eqF}; in this case, in fact, $W$ factorizes as 
$W^{\rm gm}(\delta_0;\delta_1,\dots,\d_n;S_n)=
W^{\rm gm}(\delta_0;\delta_1,\dots,\d_m;S_m)W^{\rm gm}(\delta_m;\delta_{m+1},\dots,\d_n;S_n-S_m)$.
The fact that the integral over $\d_{m-1}$ runs over
$\d_{m-1}\leq B^a_{m-1}$ while the integral over $\d_m$ runs over $\d_m\geq B^a_m$ implies that
the factor 
$(2\pi\eps)^{-1/2}\exp\{ -(\d_m-\d_{m-1})^2/(2\eps)\}$
which appears in the first $W$ factor (see \eq{W}) becomes, in the continuum limit, a Dirac delta which forces $\d_m$ to become equal to $B^a_m$ in the second factor $W^{\rm gm}$, so inside the integral we can write
\bees
&&\hspace*{-4mm}W^{\rm gm}(\delta_0;\dots,\d_n;S_n)=\\
&&\hspace*{-4mm}W^{\rm gm}(\delta_0;\dots,\d_m;S_m)W^{\rm gm}(B^a_m;\delta_{m+1},\dots,\d_n;S_n-S_m)\, .\nn
\ees
The first $W$ factor, integrated over $d\d_1,\ldots,d\d_m$ in \eq{num}, cancels  by construction
the denominator 
${\cal D}$, and the integral over $d\d_n$ in \eq{num} can be treated as 
in \eq{eq2}. As a result, we get again \eq{eqF}, except that $\d_0$ is now replaced by $B^b_m$ and $S_n$ by $S_n-S_m$.
The same argument can be repeated in the non-Markovian case. Simply,
all derivatives $\pa_i$, when acting on the  second factor
$W^{\rm gm}(\delta_m;\delta_{m+1},\ldots,\delta_n;S_n)$,  must be evaluated at $\d_m=B^a_m$ (including derivatives $\pa_i$ with $i=m$). In this way  the dependence on $\d_m$ remains only in the first $W^{\rm gm}$ factor, and the derivation of \eqst{eq1}{eq:der} goes through.

After all these considerations, we may write the numerator (\ref{num}) in general as

\bees
&&{\cal N}=\frac{\eps^2\pa^2}{\pa S'_m\pa S_n}
\int_{-\infty}^{B^a_1}d\d_1\ldots \int_{-\infty}^{B^a_m}d\d_{m}\nonumber\\
&\times&
\int_{-\infty}^{B^b_{m+1}}d\d_{m+1}\ldots \int_{-\infty}^{B^b_{n}}d\d_{n}\nonumber\\
&\times&\left[1+\hat{f}\,\right]\left.W^{\rm gm}(\d_0;\d_1,\ldots \d_m;S'_m)\right|_{S'_m=S_m}\nonumber\\
&\times&
\left.W^{\rm gm}(\d_m;\d_{m+1}\ldots ,\d_{n};S_{n}-S_m)\right|_{\delta_m=B^a_m}
\label{num2}\, .
\ees
where the differential operator $\hat{f}$, which acts on both $W$ factors, has the general form
\bees
\hat{f}&=&\sum_{i=1}^n a(S_i)\pa_i+\sum_{i,j=1}^n b(S_i,S_j)\pa_i\pa_j\nn\\
&&+
\sum_{i,j,k=1}^n c(S_i,S_j,S_k)\pa_i\pa_j\pa_k+
\ldots
\ees
and  takes into account the effects of the  the moving barrier and/or the non-Gaussian contributions. The denominator (\ref{den}), as we saw, is given by
\bees\label{den2}
&&{\cal  D}=-\eps\frac{\pa}{\pa S_m}\int_{-\infty}^{B^a_1}d\d_1\ldots \int_{-\infty}^{B^a_m}d\d_{m}\nonumber\\
&\times& \left[1+\hat{f}\,\right]
W^{\rm gm}(\d_0;\d_1,\ldots \d_m;S_m)\, .
\ees
The ratio ${\cal N}/{\cal  D}$ defining the conditional two-barrier probability is therefore proportional to $\eps$ and the flux
${\cal F}_{\rm mb}$, which is obtained dividing further by $\eps$, is therefore equal to the ratio of \eq{num2} and \eq{den2}, i.e.

\be
{\cal F}_{\rm mb}(B^b_n,S_n|B^a_m,S_m)=\lim_{\eps\rightarrow 0^+}\left(\frac{{\cal N}}{\eps\,{\cal D}}\right)\, .
\ee
Notice that in the fully Markovian case and constant spherical collapse barrier the operator $\hat{f}$ vanishes. Then factorization applies and one recovers the standard result of \cite{lc} that
the two-barrier conditional probability with constant barriers $\delta_b(z_b)$
 and $\delta_c(z_c)$ is given by the usual first-crossing rate where the initial
conditions are such that the smoothed density contrast is  $\delta_m=\delta_b$ at time $S_m$

\begin{equation}
\label{usual}
{\cal F}_{\rm sph}(\delta_c,S_n|\delta_b,S_m)=
\frac{(\delta_c-\delta_b){\rm e}^{-(\delta_c-\delta_b)^2/(2(S_n-S_m))}}
{\sqrt{2\pi}(S_n-S_m)^{3/2}}
\, .
\end{equation}

\subsection{The two-barrier conditional probability: the moving barrier case and Gaussian initial conditions}

In the case in which the barrier threshold is moving, as for the Sheth-Tormen barrier,
and the initial conditions are Gaussian, the computation of the two-barrier conditional probability is
very similar to the one we have performed in the previous section for the halo bias.
The key point is to perform, in \eqs{num}{den}, the following
shift of the integration variables $\d_i$ with $i\neq m$ and $i\neq n$
\begin{eqnarray}
\label{shift1}
&&\delta_i\rightarrow \delta_i-\sum_{p=1}^\infty\frac{B_m^{a,(p)}}{p!}
\, \left(S_i-S_m\right)^p,i=1,\cdots,m-1\,, \nonumber\\
&&\delta_i\rightarrow \delta_i-\sum_{p=1}^\infty\frac{B_n^{b,(p)}}{p!}
\, \left(S_i-S_n\right)^p,i=m+1,\cdots,n-1\,,\nonumber\\
&&
\ees
(where $B_m^{a,(p)}$ denotes the $p$-th derivative with respect to $S$ of the barrier $B^a(S)$, evaluated in $S=S_m$, and similarly $B_m^{b,(p)}$  is the  $p$-th derivative
of the barrier $B^b(S)$), while no shift is
made for the variables  $\delta_m$ and $\delta_n$.
It is easy to convice oneself that,
exploiting the factorization property (\ref{facto1}), the contributions to the two-barrier conditional probability coming from the random walks starting at $\delta_0=0$ at $S=0$ and crossing the barrier
$B_m$ for the first time at $S_m$ cancel and one is left with
\begin{eqnarray}
&&\hspace*{-8mm}
{\cal F}_{\rm mb}(B^b_n,S_n|B^a_m,S_m)
=-\frac{\pa}{\pa S_n}
\int_{-\infty}^{B^b_{m+1}}d\d_{m+1}\ldots \int_{-\infty}^{B^b_{n}}d\d_{n}\nonumber\\
&&W(B_m;\d_{m+1}\ldots ,\d_{n};S_{n}-S_m)\nonumber\\
&=&
\frac{{\rm e}^{-(B^b_n-B^a_m)^2/(2(S_n-S_m))}}{\sqrt{2\pi}(S_n-S_m)^{3/2}}
\left[B^b_n-B^a_m+{\cal P}(S_m,S_n)\right]\, .\nonumber\\
&&
\label{FcondG}
\ees
This finding coincides with  the result obtained by  \cite{ST2002}; in fact it is valid for any generic moving barrier (and therefore also for the ST barrier with $a$ different from unity)   and justifies what found in \cite{ST2002}  on more rigorous grounds.
As was the case for the unconditional mass function, the height of the barrier diverges for
$(S_n-S_m)\rightarrow\infty$, so not all trajectories intersect the second barrier. It seems reasonable (\cite{ST2002}) to associate the fraction of random walks that do not with the fraction of the parent halo mass that is not associated with bound subclumps.

\begin{figure}
\centering
\includegraphics[width=0.4\textwidth]{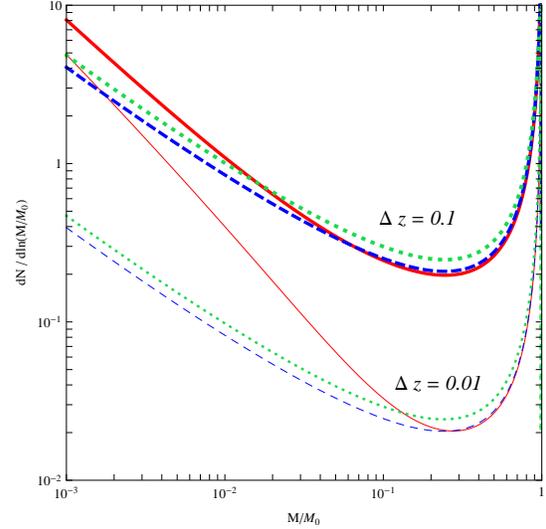}
\caption{
\label{fig:condG}
The conditional mass functions for the progenitor halos of a
descendant halo of mass $M_0=10^{15} h^{-1} M_\odot$ at $z=0$, according to
Eq.~(\ref{FcondG}) with $a=0.707$ (solid red) and to
the spherical collapse model with $a=0.707$ (dashed blue) and 
with $a=1$ (dotted green).
Two different look-back times are shown: $\Delta z=0.1$ (thick lines),
and $\Delta z= 0.01$ (thin lines).
}
\end{figure}
Once the two-barrier first-crossing rate is known, we may compute the conditional mass function,
eq.~(\ref{condmassfunction}),
to study how many progenitors at $z_b$ are associated with a descendent halo mass $M_0$ at $z_a$.
In Fig.~\ref{fig:condG} we show the conditional mass functions for the progenitor halos of a
descendant halo of mass $M_0=10^{15} h^{-1} M_\odot$ at $z_a=0$ and look-back times
$\Delta z=(z_b-z_a)=0.1$ and $0.01$. 
We plot the moving barrier result Eq.~(\ref{FcondG}) 
for the ellipsoidal barrier in eq.~(\ref{ST}) with $a=0.707$, 
and the spherical collapse result Eq.~(\ref{usual})  (both with $a=0.707$ and $a=1$).

\subsection{The two-barrier conditional probability: the moving barrier case and non-Gaussian initial conditions}
In the case in which non-Gaussian initial conditions are present, we can deal with the problem in the same way we have been treating the computation of the halo bias in the case of a non-Gaussian theory. 
Expanding for weak non-Gaussianities in the expression (\ref{num2}) brings down  the sum 

\begin{eqnarray}
\sum_{i,j,k=1}^{n}\langle\d_i\d_j\d_k\rangle_c\partial_i\partial_j\partial_k&=&
\sum_{i,j,k=1}^{m}\langle\d_i\d_j\d_k\rangle_c\partial_i\partial_j\partial_k\nonumber\\
&+&\sum_{i,j,k=m+1}^{n}\langle\d_i\d_j\d_k\rangle_c\partial_i\partial_j\partial_k\nonumber\\
&+&3\sum_{i=1}^{m}\sum_{j,k=m+1}^{n}\langle\d_i\d_j\d_k\rangle_c\partial_i\partial_j\partial_k\nonumber\\
&+&3\sum_{i,j=1}^{m}\sum_{k=m+1}^{n}\langle\d_i\d_j\d_k\rangle_c\partial_i\partial_j\partial_k\, .
\nonumber\\
&&
\end{eqnarray}
We now perform   the Taylor expansion (\ref{pqrs}) 
and retain only the leading terms

\begin{eqnarray}
\sum_{i,j,k=1}^{n}\langle\d_i\d_j\d_k\rangle_c\partial_i\partial_j\partial_k&\simeq&
\sum_{i,j,k=1}^{m}\langle\d_m^3\rangle_c\partial_i\partial_j\partial_k\nonumber\\
&+&\sum_{i,j,k=m+1}^{n}\langle\d_n^3\rangle_c\partial_i\partial_j\partial_k\nonumber\\
&+&3\sum_{i=1}^{m}\sum_{j,k=m+1}^{n}\langle\d_m\d_n^2\rangle_c\partial_i\partial_j\partial_k\nonumber\\
&+&3\sum_{i,j=1}^{m}\sum_{k=m+1}^{n}\langle\d^2_m\d_n\rangle_c\partial_i\partial_j\partial_k\, .\nonumber\\
&&
\end{eqnarray}
As $B_m^a<B_n^b$ we may further approximate this sum ignoring the terms 
proportional to $\langle\d_m^3\rangle_c$ and $\langle\d_m^2\delta_n\rangle_c$
with respect to the others

\begin{eqnarray}
\sum_{i,j,k=1}^{n}\langle\d_i\d_j\d_k\rangle_c\partial_i\partial_j\partial_k&\simeq&
\sum_{i,j,k=m+1}^{n}\langle\d_n^3\rangle_c\partial_i\partial_j\partial_k\nonumber\\
&+&3\sum_{i=1}^{m}\sum_{j,k=m+1}^{n}\langle\d_m\d_n^2\rangle_c\partial_i\partial_j\partial_k\, .
\nonumber\\
\end{eqnarray}
Again, the 
crucial point is that, when considering a moving barrier,  the shift of variables (\ref{shift1}) does not involve neither $\delta_m$ nor $\delta_n$. 
This allows  factorization and cancellations between the numerator (\ref{num2}) and 
the denominator (\ref{den2}).
The non-Gaussian two-barrier conditional probability ${\cal F}_{\rm mb\,NG}$ then becomes
\begin{eqnarray}
&&\hspace*{-8mm}{\cal F}_{\rm mb\,NG}(B^b_n,S_n|B^a_m,S_m)=\nn\\
&&\hspace*{-8mm}
-\frac{\pa}{\pa S_n }\int_{-\infty}^{B^b_n}d\d_n\, \Pi_{\rm two} (\delta_n,S_n|  B^a_m,S_m)\nonumber\\
&&\hspace{-8mm}
+\frac{1}{6}\frac{\pa^4}{\pa S_n\pa (B^b_n)^3}\left(\langle\d_n^3\rangle_c\int_{-\infty}^{B^b_n}d\d_n\, \Pi_{\rm two} ( \delta_n,S_n|  B^a_m, S_m)\right)\nonumber\\
&&\hspace*{-8mm}
-\frac{3}{6}\hspace*{-0.5mm}\frac{\pa^4}{\pa S'_m\pa S_n\pa (B_n^b)^2}\left(\hspace*{-2mm}
\left.\langle\d_m\d_n^2\rangle_c\hspace*{-1mm}\int_{-\infty}^{B^a_m}\hspace*{-1mm}d\d_m
\frac{\partial\Pi_{\rm mb}(\d_m; S'_m)}{\partial B^a_m}\right|_{S'_m=S_m}\right.
\nonumber\\
&&\hspace*{-8mm}\times
\left.\int_{-\infty}^{B^b_n}d\d_n\, \Pi_{\rm two} ( \delta_n,S_n|  B^a_m, S_m)\right)\frac{1}{{\cal F}_{\rm mb}(S_m)}
\nonumber\\
&&\hspace*{-8mm}-\frac{3}{6}\frac{\pa^4}{\pa S'_m\pa S_n\pa (B_n^b)^2}\left(\hspace*{-2mm}
\left.\langle\d_m\d_n^2\rangle_c\int_{-\infty}^{B^a_m}d\d_m\,\Pi_{\rm mb}(\d_m;B^a_m,S'_m)\right|_{S'_m=S_m}\right.
\nonumber\\
&&\hspace*{-8mm}\times
\left.\int_{-\infty}^{B^b_n}d\d_n\partial_m \left.\Pi_{\rm two} ( \delta_n,S_n| \d_m, S_m)\right|_{\delta_m=B^a_m}\right)\frac{1}{{\cal F}_{\rm mb}(S_m)}\nn\\
&&\hspace*{-5mm}\equiv 
{\cal F}^{\rm (I)}_{\rm mb\,NG}
+{\cal F}^{\rm (II)}_{\rm mb\,NG}
+{\cal F}^{\rm (III)}_{\rm mb\,NG}
+{\cal F}^{\rm (IV)}_{\rm mb\,NG}\,,
\label{fmbngcond}
\ees
where $\Pi_{\rm mb}$ can be read from eq.~(\ref{aaaaa}), ${\cal F}_{\rm mb}(S_m)$ from eq.~(\ref{aaprime}) and 
\begin{eqnarray}
\label{pitwo}
&&\Pi_{\rm two} ( \delta_n, S_n| \d_m, S_m) =\nonumber\\
&&\frac{1}{\sqrt{2\pi (S_n-S_m)}}\,
\left(  {\rm e}^{-(\d_n-\d_m)^2/(2(S_n-S_m))}\right.\nonumber\\
&-& \left.{\rm e}^{-(2B^b_n-\d_n-\d_m)^2/(2(S_n-S_m))} \right)\nonumber\\
&+&{2(B^b_n-\delta_n)\over \sqrt{2\pi} (S_n-S_m)^{3/2}}
 {\rm e}^{-(2B^b_n-\delta_n-\d_m)^2/(2 (S_n-S_m))}{\cal P}_{mn}\nn\\
&-&{2(B^b_n-\delta_n)^2\over \sqrt{2\pi} (S_n-S_m)^{5/2}}  
 {\rm e}^{-(2B^b_n-\delta_n-\d_m)^2/(2 (S_n-S_m))}
{\cal P}_{mn}^2\,.
\nonumber\\
&&
\end{eqnarray}
Notice that repeating with care the steps described in subsection (\ref{preliminiaries}),
in eq.~(\ref{fmbngcond}) we do not have to differentiate the cumulant 
$\langle \d_m\d_n^2\rangle_c$ with respect to $S'_m$ and $S_n$.

Let us set $\langle\d_n^2\d_m\rangle_c={\cal S}_{2,1}(S_n,S_m)S_n S_m^{1/2}$, where
${\cal S}_{2,1}(S_n,S_m)$ is slowly changing with $S_n$ and $S_m$ (see App.~\ref{app:fits} for
further details and useful fitting functions). 
Then, the first two terms of (\ref{fmbngcond}) read
\begin{eqnarray}
&&\hspace*{-5mm}
{\cal F}^{\rm (I)}_{\rm mb\,NG}(B^b_n, S_n| B^a_m, S_m)
+{\cal F}^{\rm (II)}_{\rm mb\,NG}( B^b_n, S_n| B^a_m, S_m)\nn\\
&&\hspace*{-5mm}=
\frac{(B^b_n-B^a_m+{\cal P}_{mn}) {\rm e}^{-{(B^b_n-B^a_m)^2\over 2(S_n-S_m)}}}
{\sqrt{2\pi}(S_n-S_m)^{3/2}}
\nonumber\\
&&\hspace*{-5mm}+\frac{S_n{\cal S}_3(S_n)}
{6\sqrt{2\pi }(S_n-S_m)^{9/2}}{\rm e}^{-{(B^b_n-B^a_m)^2\over 2(S_n-S_m)}}
\left[
(B^b_n-B^a_m)^4 S_n\right.\nn\\
&&\hspace*{-5mm}
-(B_n^b-B_m^a)^3 S_n ({\cal P}_{mn}+2 B'^b_n (S_n-S_m))
\nn\\
&&\hspace*{-5mm}
+ 2(B^b_n-B^a_m)^2(S_n{\cal P}_{mn}^2+S_n(S_n-S_m){\cal P}_{mn}B'^b_n \nn\\
&&\hspace*{-5mm}
- S_n^2+2S_m^2 -S_nS_m)\nn\\
&&\hspace*{-5mm}
+(B^b_n-B^a_m)(S_n-S_m) \left({\cal P}_{mn}(S_n+4S_m)
\right.\nn\\
&&\hspace*{-5mm}
\left.
+6S_n(S_n-S_m) B'^b_n
-4 {\cal P}_{mn}^2 S_n B'^b_n- 2S_n(S_n-S_m){\cal P}_{mn}'\right)\nn\\
&&\hspace*{-5mm}
-2(S_n-S_m){\cal P}_{mn}({\cal P}_{mn}(S_n+4S_m)+(S_n-S_m)S_n B'^b_n
\nonumber\\
&&\hspace*{-5mm}
\left.
- 4S_n(S_n-S_m){\cal P}_{mn}')
-S_n^3+4S_m^3+6S_n^2S_m-9S_n S_m^2
\right]\nn\\
&&\hspace*{-5mm}
+\frac{S_n^2{\cal S}'_3(S_n)}{3\sqrt{2\pi }(S_n-S_m)^{5/2}}
{\rm e}^{-{(B^b_n-B^a_m)^2\over 2(S_n-S_m)}}
\left[ (B^b_n-B^a_m)^2  \right.\nn\\
&&\hspace*{-5mm}
\left. - (B^b_n-B^a_m){\cal P}_{mn}
-(S_n-S_m)+2{\cal P}_{mn}^2\right] \,.
\label{termini12}
\end{eqnarray}
where the prime stands for derivative with respect to $S_n$.
As for the third and fourth terms  of Eq.~(\ref{fmbngcond}), they are
\begin{eqnarray}
&&\hspace*{-8mm}
{\cal F}^{\rm (III)}_{\rm mb\,NG}( B^b_n, S_n| B^a_m, S_m)=
-{3\over 6} 
{\cal S}_{2,1}(S_n,S_m) S_n S_m^{1/2}\nn\\
&&\times
{\partial^2\over \partial (B_n^b)^2}{\cal F}_{\rm mb}( B_n^b, S_n|B_m^a, S_m)
{\partial\over \partial B_m^a}\ln {\cal F}_{\rm mb}(S_m)
\,,
\nn\\
\label{termine3}
\end{eqnarray}
where is ${\cal F}_{\rm mb}(S_m)$ is given by eq.~(\ref{aaprime})
and
\begin{eqnarray}
&&\hspace*{-8mm}
{\cal F}^{\rm (IV)}_{\rm mb\,NG}( B^b_n, S_n| B^a_m, S_m)=
-{3\over 6} 
{\cal S}_{2,1}(S_n,S_m) S_n S_m^{1/2} \nn\\
&&\times
{\partial^3\over \partial (B_n^b)^2\partial B_m^a}{\cal F}_{\rm mb}( B_n^b, S_n|B_m^a, S_m)\,.
\label{termine4}
\end{eqnarray}

\begin{figure}
\centering
\includegraphics[width=0.4\textwidth]{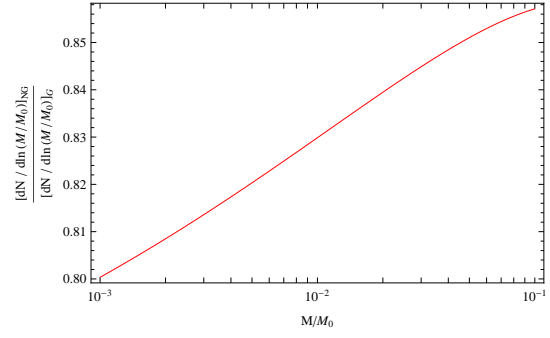}
\caption{
\label{fig:condNG}
Ratio of the conditional mass functions with and without NG, for the progenitor halos of a
descendant halo of mass $M_0= 10^{15} h^{-1} M_\odot$ at $z=0$, 
look-back time $\Delta z=0.3$, $f_{\rm NL}=50$ and $a=0.707$
}
\end{figure}
Notice that the total NG conditional probability is no longer a function of the
variables $(S_n-S_m)$ and $(B_n^b-B_m^a)$  as in the Markovian case.
In Fig. \ref{fig:condNG} we show the ratio of the conditional mass functions with and without NG, for the progenitor halos of a
descendant halo of mass $M_0=10^{15} h^{-1} M_\odot$ at $z_a=0$, 
look-back time $\Delta z=0.3$, $f_{\rm NL}=50$ and $a=0.707$.

\section{Halo formation time probability with non-Gaussianities}   
 \label{sect:formation}
 
In this section we present the results for  the probability distribution of halo  formation redshifts with the inclusion
of non-Gaussianities.
 
We follow the convention to define the  epoch of formation of a halo as the time 
 when the halo contains half of its final mass, although the generalization to an arbitrary
 fraction between  $1/2$ and $1$ is straightforward.
Let us fix the mass $M_0$ and the redshift $z_a$ of the descendent halo.
Then, the probability that  such halo had a progenitor at redshift $z_b>z_a$ 
with mass between $M_0/2$ and $M_0$ (or, equivalently, the probability
that the formation redshift is bigger than $z_b$)
 is given by eq.~(\ref{condmassfunction})
 integrated from  $M_0/2$ to $M_0$:
 \begin{eqnarray}
P(z_b;M_0, z_a)&=&\int_{S_0}^{S_h} dS_n {M_0\over M(S_n)}\nn\\
&&
\times 
{\cal F}_{\rm mb}( B(z_b; S_n), S_n | B(z_a; S_0),  S_0) \,,
\label{Pformationtime}
 \end{eqnarray}
 where $S_0\equiv S(M_0), S_h\equiv S(M_0/2)$.
From this, it is possible to find the probability distribution that the halo of mass $M_0$ at $z_a$ 
would have formed between $z_b$ and $z_b+dz_b$:
\be
p(z_b)dz_b=\left\vert{dP(z_b; M_0, z_a)\over dz_b}\right\vert dz_b\,.
\ee

In the simple case of constant barrier $\delta_c(z)$ and Gaussian initial conditions, 
the distribution of halo formation redshifts is simply given by
(see \cite{lc})
\be
p(z_b)=2\omega(z_b){\textrm{Erfc}}\left[{\omega(z_b)\over  \sqrt{2}}\right]
{d\omega(z_b)\over d z_b}\, ,
\ee
where 
\be
\omega(z_b)\equiv{\delta_c(z_b)-\delta_c(z_a)\over\sqrt{S(M_0/2)-S(M_0)}}\,,
\ee
if  it is assumed a white-noise power spectrum,  leading to $S(M)\propto M^{-1}$ (\cite{lc}).

Now let us introduce the  contribution of non-Gaussianities. The two barrier
conditional probability is given by eq.~(\ref{fmbngcond}). In order to simplify the calculation and reach a compact result we make the following
assumptions: we  ignore the cumulant $\langle\delta_m\delta_n^2\rangle$,
 keeping only the leading one $\langle\delta_n^3\rangle$ 
 (which means we only retain the terms in eq.~(\ref{termini12})); 
we consider 
 ${\cal S}_3$ as a constant and we assume that $S(M)$ scales like $M^{-1}$,
In these approximations,
the (normalized) probability distribution of formation redshifts in presence of non-Gaussianities
becomes
\begin{eqnarray}
p_{{\rm NG}}(z_b)&=&\left\{2\omega(z_b){\textrm{Erfc}}\left[{\omega(z_b)\over \sqrt{2}}\right]\right.\nn\\
&&\hspace{-1cm}
+
{4\over 3\sqrt{\pi}}{\cal S}_3  \sqrt{S(M_0)}\left[
\sqrt{2}\omega(z_b) \left(\omega^2(z_b)-3\right)
{\rm e}^{-{\omega(z_b)^2\over 2}}\right.\nn\\
&&\hspace{-1cm}
\left.\left.
+\sqrt{\pi}\left(1-\omega^2(z_b)\right)\textrm{Erfc}\left[{\omega(z_b)\over \sqrt{2}}\right]
\right]\right\}\nn\\
&&\hspace{-1cm}
\times \left[
1-{8\over 9}\sqrt{2\over \pi}{\cal S}_3 \sqrt{S(M_0)}
\right]^{-1}
{d\omega(z_b)\over d z_b}\,.
\label{az}
\end{eqnarray}
This expression is also valid for a spherical collapse model with barrier $\sqrt{a}\d_c(z)$ which
may be considered as an approximation to the ellipsoidal model  (\cite{Giocoli2007}).
In order to gain  an intuition of the impact of the non-Gaussian correction, 
for $M_0=10^{15} M_\odot h^{-1}$ and $f_{\rm NL}=50$, 
in the regions of $z_b$ where the distribution is greater than 0.1, the NG term contributes typically less than 10\%
with respect to the Gaussian one.

In Fig.~\ref{fig:formation} we show  the probability distributions of formation
redshifts of a halo of mass $M_0= 10^{15} h^{-1} M_\odot$ at $z_a=0$, both for the spherical collapse model
with constant barrier $\sqrt{a}\delta_c$ and for the ellipsoidal collapse model with barrier $B_{\rm ST}$ in eq.~(\ref{ST}). These results are obtained by integrating numerically eq.~(\ref{Pformationtime})
using the first-crossing rates in eq.~(\ref{FcondG}) and in eqs.~(\ref{termini12})-(\ref{termine4})
for Gaussian and non-Gaussian initial conditions, respectively.
The inclusion of non-Gaussianities tends to shift slightly the distributions towards higher redshifts.
Furthermore, the results for the
spherical collapse with barrier $\sqrt{a}\delta_c$ and ellipsoidal barrier are quite close to each other, confirming the suggestions of \cite{Giocoli2007}. 

The mass dependence of the variance $S(M)$ enters into eq.~(\ref{Pformationtime}) in an important way.
We make use of the numerical fit (\ref{SofMfit}) (solid lines in Fig.~\ref{fig:formation}), but we have verified 
that using a different fit to $S(M)$, like the one of \cite{Neistein}, does not change the solid curves appreciably. 
For comparison, we also show the distributions (dashed lines) one would obtain by using the simple scaling
$S(M)\propto M^{-1}$, which is very useful to carry out analytical calculations and it is commonly used
in the literature. However, as shown in the figure, the use of this simple scaling leads to 
non-negligible differences with respect to a more accurate numerical fit. Note also that for $z_b\rightarrow z_a$ the numerical fit $S(M)$ gives values of $dp/d z_b$ systematically higher than
for the case $S(M)\propto M^{-1}$. This point was already noted in \cite{lc} where it was stressed that using a variance with a scaling different from $M^{-1}$ gives rise to probabilities $dp/d z_b$ which
do not vanish for $z_b\rightarrow z_a$. This is probably due to the fact that the particles
that the analysis of trajectories tags as having a given mass $M$ are not really grouped into objects of mass $M$ (\cite{lc}).

The analytical prediction (\ref{az}), which has been found  for a constant barrier under the assumption  $S(M)\propto M^{-1}$ and leading NG term,  turns out to be in very good agreement with the numerical result obtained by integrating eq.~(\ref{Pformationtime}) numerically, under the same assumption for $S(M)$.

\begin{figure}
\centering
\includegraphics[width=0.4\textwidth]{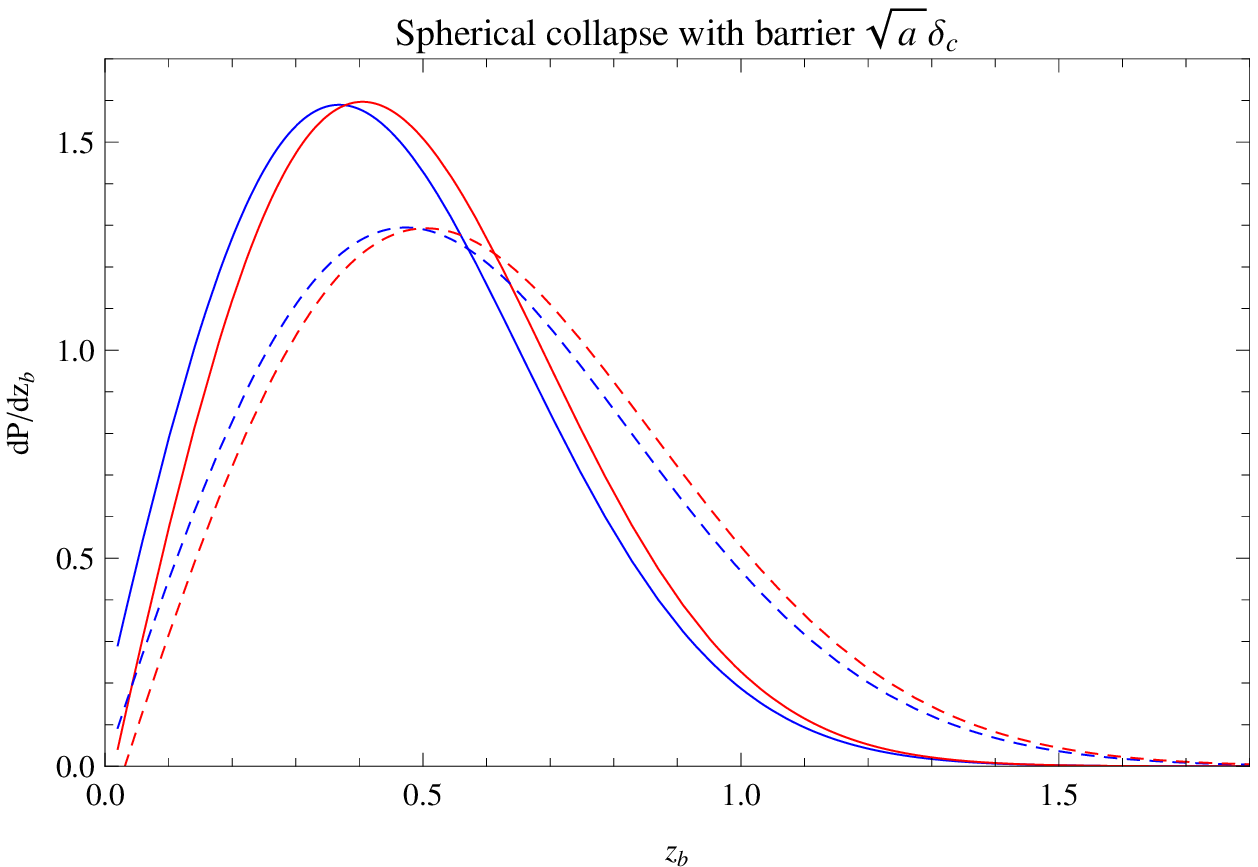}
\includegraphics[width=0.4\textwidth]{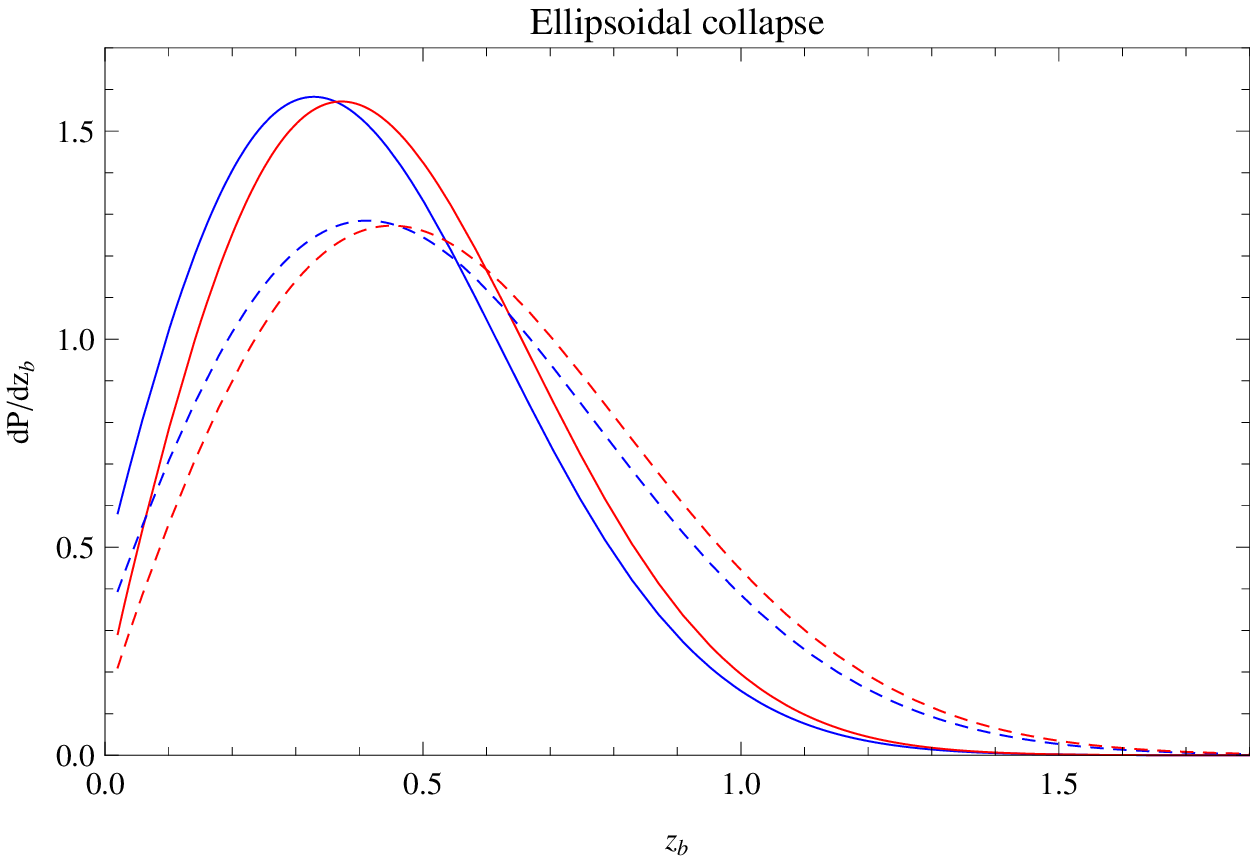}
\caption{
\label{fig:formation}
Probability distributions for the formation redshift $z_b$ of a halo of
 mass $M_0= 10^{15} h^{-1} M_\odot$ at $z_a=0$, in the spherical collapse model
 with barrier $\sqrt{a}\delta_c$ (top panel)  and in the  ellipsoidal collapse model  (bottom panel).
The Gaussian case is shown in blue, while the 
 non-Gaussian one,  with $f_{\rm NL}=50$, is in red.
Solid lines correspond to the numerical fit to $S(M)$ in eq.~(\ref{SofMfit}) while for the 
dashed lines the simple approximation $S(M)\propto M^{-1}$ is used.
}
\end{figure}

\section{Conclusions}
\label{sect:conclusions}

In the excursion set theory the  density perturbations depend stochastically with the smoothing scale
and the  computation of the halo  mass function is mapped into   the so-called first-passage time problem in the presence of a barrier.  Other properties of dark matter halos,  such as   halo bias, accretion 
rate, formation time,  merging rate and the formation history of halos, can be studied using the excursion set formalism by 
computing   the conditional probability
with non trivial initial conditions and the conditional two-barrier
crossing rate. 

 In this paper we have performed 
the calculations of such conditional probabilities   
in the presence of a generic moving barrier and  for  both   Gaussian and non-Gaussian initial conditions. 
 Our generic results can therefore be applied to the case of the ellipsoidal
 collapse where the barrier is moving, given by the expression (\ref{ST}), and to the case of the diffusive barrier discussed in MR2.  Our findings  include
  the non-Markovianity  of the random walks induced by NG.
  
Let us summarize the main results of this paper:
\begin{itemize}%
\item  assuming a sharp filter in momentum space, the first two halo bias coefficients for a generic moving barrier  
(eqs.~(\ref{d1haloG}) and (\ref{d2haloG})) and their corrections due to 
non-Gaussianities (eqs.~(\ref{d1haloNG}) and (\ref{d2haloNG}));\\
\item the conditional mass function (eq.~(\ref{condmassfunction}))
for a generic moving barrier with Gaussian initial conditions
(eq.~(\ref{FcondG})) which reproduces what found in \cite{ST2002} and in presence of non-Gaussianities  
(eqs.~(\ref{termini12}), (\ref{termine3}), (\ref{termine4})), again for a sharp filter in momentum
space; \\
\item the probability distribution of the halo formation time, including non-Gaussianities; 
we have provided numerical results for the spherical and the ellipsoidal collapse models (Fig.~\ref{fig:formation})
and an analytical approximation (eq.~(\ref{az})) valid for constant barrier,  again for a sharp filter in momentum
space.
\end{itemize}

Our calculations have revealed a different scaling of the Gaussian halo bias parameter at high halo masses from what found in \cite{SMT}, our prediction being about 20\% higher. This seems to
go in the right direction to fit better the N-body data by \cite{Tinker2010} when such a comparison
is possible. 

As an application of our findings, it would be interesting to investigate  e.g. the NG halo assembly bias,
recently  discussed in \cite{STassembly, Gao, Reid}.

\vspace{5mm}
\vspace{5mm}
\noindent
\paragraph*{Acknowledgements.}
We thank P.S. Corasaniti for correspondence
and A. Paranjape, T.Y. Lam and R.K. Sheth
for pointing out a typo in the first manuscript.
The work of ADS is supported by the Fonds National Suisse under 
contract 200021-125237.
The work of MM is supported by the Fonds National Suisse.
The work of AR is supported by 
the European Community's Research Training Networks 
under contract MRTN-CT-2006-035505.
\vspace{5mm}

\appendix

\section{Numerical fits to cumulants}
\label{app:fits}

We report here a collection of numerical results we have found and used throughout the 
paper. For the window function, we have assumed  a top-hat in wavenumber space, 
but with a mass-to-smoothing scale relation of a real space top-hat filter. 
The following cosmological parameters from  \cite{WMAP7} are used:
$h=0.703, \Omega_\Lambda=0.729, \Omega_m=0.271, \Omega_b=0.0451, \sigma_8=0.809$.
 
The variance $S\equiv \langle\delta_R^2\rangle$ depends on  the smoothing scale $R$
and, in turn, on the halo mass $M$ (in units of $M_\odot h^{-1}$) according to 
 the fitting function
\begin{eqnarray}
S(M)&=&c_0 \left[1+{c_1\over 10} M^{1/10}
+{c_2\over 10^2} M^{2/15}+ 
\right.\nn\\
&&\left. {c_3\over 10^3} M^{1/6} + {c_4\over 10^4} M^{1/5}
\right]^{-10}
\label{SofMfit}
\end{eqnarray}
with $c_0=7.2\times 10^2, c_1=1.2, c_2= -5.8 , c_3=9.5, 
c_4= 2.6 $,
which agrees rather well with the one reported in \cite{Neistein}, who instead make
use of a top-hat filter in real space.

For the  scale-dependence of ${\cal S}_3\equiv \langle\delta^3(S)\rangle/ S^2$ we have found the following simple fitting
formula, computed along the lines of \cite{MVJ},  with updated cosmological parameters:
\be
{\cal S}_3(S)\simeq {2.9\times 10^{-4}\over S^{0.3}} f_{\rm NL}\,.
\ee
The time-variation of ${\cal S}_3$ is such that $S {\cal S}'_3/{\cal S}_3\sim 0.3$.
Alternatively, one may define the quantity 
\be
\label{epsilon1}
\epsilon_1(S)\equiv {\langle\delta^3(S)\rangle\over S^{3/2}}\,,
\ee
which varies more slowly in $S$, since $S \epsilon'_1/\epsilon_1\sim 0.2$.
This quantity is well-fitted by the formula
\be
\epsilon_1(S)\simeq 2.9\times 10^{-4} \, S^{0.2} f_{\rm NL}\,.
\ee

For the cumulant  $ \langle\delta^2(S_1)\delta(S_2)\rangle$, we have found that it scales approximately
like $S_1\sqrt{S_2}$. Therefore, it is convenient to define a slowly-varying ${\cal S}_{2,1}(S_1,S_2)$
as
\be
{\cal S}_{2,1}(S_1,S_2)={\langle\delta^2(S_1)\delta(S_2)\rangle
\over S_1\sqrt{S_2}}\,.
\ee
Having fixed $S_2\equiv S_{15}\simeq 0.2 $, the variance corresponding to a halo mass of 
$10^{15} M_\odot h^{-1}$,
we have found the fitting formula (for $S>S_{15}$)
\be
{\cal S}_{2,1}( S, S_{15})\simeq {2.4 \times 10^{-4}\over S^{0.02}} f_{\rm NL}
\,.
\ee
Its very mild dependence on $S$ justifies the assumption made in the text, where
we consider it a constant.

\section{The two-barrier first crossing rate with a top-hat window function in real space}
\label{app:realspace}

\noindent
In this Appendix we would like to extend the computation of the  two-barrier first crossing rate
to the case in which the window function for the smoothed density contrast is a top-hat in real space.
An analogous computation has been perfomed by \cite{ma} for the halo bias and we are going to use
many of the results of the Appendix of that paper. The choice of a top-hat filter in real space
introduces by itself a level of non-Markovianity. The latter is manifest in the  
two-point correlator of the smoothed density contrast (MR1)
\be\label{rever} 
   \langle\delta (S_i)\delta (S_j)\rangle = {\rm
  min}(S_i,S_j) + \D(S_i,S_j)\, , 
\ee 
where $\D(S_i,S_j)=\D(S_j,S_i)$ and, for $S_i\leq S_j$, the function
$\D(S_i,S_j)$ is well approximated by
\be \label{approxDelta}
\D(S_i,S_j)\equiv \D_{ij}\simeq \kappa\, \frac{S_i(S_j-S_i)}{S_j}\, ,\,\,S_i<S_j\, ,
\ee
with $\kappa\approx 0.44\,\,(0.35)$ 
for a tophat (Gaussian) filter in coordinate space. The parameter $\kappa$ gives a
measure of the non-Markovianity of the stochastic process. 

We perform the computation of the two-barrier crossing rate assuming a spherical collapse and we will
extend them to the diffusive barrier model introduced by \cite{MR2} at the end. The barriers
are called $\delta_b$ and $\delta_c$.

To perform the computation we use the technique discussed in detail in 
 MR1. We consider first the numerator in \eq{Pbias}.
 The first step is to express the non-Markovian  $W$ in terms of $W^{\rm gm}$,
\bees\label{WnNG}
&&W(\delta_0;\ldots ,\delta_n;S_n)=\nn\\
&&\Dl\, \,
e^{ i\sum_{i=1}^n\lambda_i\delta_i
-\frac{1}{2}\sum_{i,j=1}^n\lambda_i\lambda_j
({\rm min}(S_i,S_j) + \D_{ij})}\nn\\
&&\simeq
W^{\rm gm} (\delta_0;\ldots ,\delta_n;S_n)\nn\\
&&+\frac{1}{2}\sum_{i,j=1}^n\D_{ij}\pa_i\pa_j
W^{\rm gm} (\delta_0;\ldots ,\delta_n;S_n)\, .
\ees
As usual it  is convenient to split the sum into various pieces
 \begin{eqnarray}
 \label{sumsum}
\frac{1}{2}\sum_{i,j=1}^{n}\Delta_{ij}\partial_i\partial_j&=&
\frac{1}{2}\sum_{i,j=1}^{m-1}\Delta_{ij}\partial_i\partial_j+
\sum_{i=1}^{m-1}\Delta_{im}\partial_i\partial_m\nn\\
&&\hspace*{-4mm}+
\frac{1}{2}\sum_{i,j=m+1}^{n-1}\Delta_{ij}\partial_i\partial_j+
\sum_{i=m+1}^{n-1}\Delta_{in}\partial_i\partial_n\nn\\
&&\hspace*{-4mm}+\sum_{i=1}^{m-1}\Delta_{in}\partial_i\partial_n+
\Delta_{mn}\partial_m\partial_n\nn\\
&&\hspace*{-4mm}+\sum_{i=1}^{m-1}\sum_{j=m+1}^{n-1}\Delta_{ij}\partial_i\partial_j+
\sum_{j=m+1}^{n-1}\Delta_{mj}\partial_j\partial_m\, ,\nn\label{A3}\\
\end{eqnarray}
where when sums are from 1 to $m$ the non-Markovian kernel has to be thought as a functio of $S'_m$.
The goal is to compute the numerator (\ref{num2}) and the denominator (\ref{den2}).
Consider first the contribution from the first line of (\ref{sumsum}). Its contribution to 
the numerator in (\ref{num2}) (a part from the time differentation with respect to
$S_m'$ and $S_n$) can be written as 
\bees
&& \int_{-\infty}^{\delta_b}d\delta_1\cdots  \int_{-\infty}^{\delta_b}d\d_{m}
 \int_{-\infty}^{\delta_c}d\d_{m+1}\cdots \int_{-\infty}^{\delta_c} d\delta_{n}\nn \\
&&\[\frac{1}{2}\sum_{i,j=1}^{m-1}\Delta_{ij}\partial_i\partial_j+
\sum_{i=1}^{m-1}\Delta_{im}\partial_i\partial_m\]\nn\\
&&\hspace*{-8mm}\times W^{\rm gm}(\delta_0;\ldots, \delta_m;S'_m)
W^{\rm gm}(\delta_m;  \ldots ,\delta_n; S_n-S_m)\nn\\
&=&\hspace*{-3mm}
 \int_{-\infty}^{\delta_b}d\delta_1\cdots d\delta_{m}
\[\frac{1}{2}\sum_{i,j=1}^{m-1}\Delta_{ij}\partial_i\partial_j+
\sum_{i=1}^{m-1}\Delta_{im}\partial_i\partial_m\]\nn\\
&&\hspace*{-4mm}\times W^{\rm gm}(\delta_0;\ldots, \delta_m;S'_m)\nn\\
&&\hspace*{-4mm}\times\int_{-\infty}^{\delta_c}d\d_{m+1}\cdots d\delta_{n}
W^{\rm gm}(\delta_m;  \ldots ,\delta_n; S_n-S_m)\nn \\
&+&\int_{-\infty}^{\delta_b}\hspace*{-2mm}d\delta_1\cdots d\d_{m}
\sum_{i=1}^{m-1}\Delta_{im}\partial_i
W^{\rm gm}(\delta_0;\ldots, \delta_m;S'_m)\nn\\
&&\hspace*{-4mm}\times\int_{-\infty}^{\delta_c}
\hspace*{-2mm}d\d_{m+1}\cdots d\delta_{n}
\pa_mW^{\rm gm}(\delta_m;  \ldots ,\delta_n; S_n-S_m)\, .
\label{A4}
\ees
The first term is easily dealt with by observing that
\bees
&&\int_{-\infty}^{\delta_c}d\d_{m+1}\cdots d\delta_{n}
W^{\rm gm}(\delta_m;  \ldots ,\delta_n; S_n-S_m)\nn\\
&&=\int_{-\infty}^{\delta_c}d\d_{n}
\Pi^{\rm gm}(\d_m;\d_n;S_n-S_m)\, .
\ees
Combining this with the contribution coming from the 
zero-th order term $W^{\rm gm} (\delta_0;\ldots ,\delta_n;S_n) $ in 
\eq {WnNG} and using again the factorization property  of $W^{\rm gm}$, we therefore get
\bees
&&\int_{-\infty}^{\delta_c}d\d_{n}\Pi^{\rm gm}(\d_m;\d_n;S_n-S_m)
\int_{-\infty}^{\delta_b}d\delta_1\cdots d\d_{m}\nn\\
&&\times \[1+\frac{1}{2}\sum_{i,j=1}^{m-1}\Delta_{ij}\partial_i\partial_j+
\sum_{i=1}^{m-1}\Delta_{im}\partial_i\partial_m\]\nn\\
&&\times W^{\rm gm}(\delta_0;\ldots, \delta_m;S'_m)\nn\\
&+&\int_{-\infty}^{\delta_b}\hspace*{-2mm}d\delta_1\cdots d\d_{m}
\sum_{i=1}^{m-1}\Delta_{im}\partial_i
W^{\rm gm}(\delta_0;\ldots, \delta_m;S'_m)\nn\\
&&\hspace*{-4mm}\times\int_{-\infty}^{\delta_c}
\hspace*{-2mm}d\d_{m+1}\cdots d\delta_{n}
\pa_mW^{\rm gm}(\delta_m;  \ldots ,\delta_n; S_n-S_m)\, ,
\ees
where one has to recall that the derivatives $\partial_m$ acting on $W^{\rm gm}(\delta_m;  \ldots ,\delta_n; S_n-S_m)$ have to be evaluted at $\delta_m=\delta_c$. Notice that the term in brackets
gives just the expansion to ${\cal O}(\kappa)$ of the denominator (\ref{den2}) and therefore it will provide
the usual Markovian tow-barrier crossing rate (\ref{usual}). The other terms contribute to the 
numerator (\ref{num2}) as
\be
\left.\frac{\partial^2}{\partial S_m'\partial S_n}\int_{-\infty}^{\delta_b}\hspace*{-2mm}d\delta_m
\int_{-\infty}^{\delta_c}\hspace*{-2mm}d\delta_n(N_a+N_b+N_c+N_d)\right|_{S_m'=S_m},
\ee
where 
\begin{eqnarray}
N_a&=&\int_{-\infty}^{\delta_b}\hspace*{-2mm}d\delta_1\cdots d\d_{m-1}
\sum_{i=1}^{m-1}\Delta_{im}\partial_i
W^{\rm gm}(\delta_0;\ldots, \delta_m;S'_m)\nn\\
&&\hspace*{-4mm}\times\int_{-\infty}^{\delta_c}
\hspace*{-2mm}d\d_{m+1}\cdots d\delta_{n-1}\left.
\pa_mW^{\rm gm}(\delta_m;  \ldots ,\delta_n; S_n-S_m)\right|_{\d_m=\d_b}\, ,\nn\\
\ees
\bees
N_b&=&\int_{-\infty}^{\delta_b}\hspace*{-2mm}d\delta_1\cdots d\d_{m-1}
\int_{-\infty}^{\delta_c}\hspace*{-2mm}d\d_{m+1}\cdots d\delta_{n-1}\label{Nb0}\\
&&\[
\frac{1}{2}\sum_{i,j=m+1}^{n-1}\Delta_{ij}\partial_i\partial_j+
\sum_{i=m+1}^{n-1}\Delta_{in}\partial_i\partial_n\]\nn\\
&&\hspace*{-8mm}\times W^{\rm gm}(\delta_0;\ldots, \delta_m;S'_m)
W^{\rm gm}(\delta_b;  \ldots ,\delta_n; S_n-S_m)\, ,\nn
\ees
\bees
N_c&=&\int_{-\infty}^{\delta_b}\hspace*{-2mm}d\delta_1\cdots d\d_{m-1}\int_{-\infty}^{\delta_c}\hspace*{-2mm}
d\d_{m+1}\cdots d\delta_{n-1}\\
&&\[
\sum_{i=1}^{m-1}\Delta_{in}\partial_i\partial_n+
\Delta_{mn}\partial_m\partial_n
\]\nn\\
&&\hspace*{-8mm}\times W^{\rm gm}(\delta_0;\ldots, \delta_m;S'_m)
W^{\rm gm}(\delta_b;  \ldots ,\delta_n; S_n-S_m)\, ,\nn
\ees
\bees
N_d&=&\int_{-\infty}^{\delta_b}\hspace*{-2mm}d\delta_1\cdots d\d_{m-1}
\int_{-\infty}^{\delta_c}\hspace*{-2mm}d\d_{m+1}\cdots d\delta_{n-1}\nn\\
&&\left[
\sum_{i=1}^{m-1}\sum_{j=m+1}^{n-1}\Delta_{ij}\partial_i\partial_j+
\sum_{j=m+1}^{n-1}\Delta_{mj}\partial_j\partial_m
\right]\nn\\
&&\hspace*{-8mm}\times W^{\rm gm}(\delta_0;\ldots, \delta_m;S'_m)
W^{\rm gm}(\delta_b;  \ldots ,\delta_n; S_n-S_m)\, ,\nn\\
&&
\label{NDD}
\ees
Notice that  in the last term $N_d$  particular attention has to be paid on how to
reconstruct the derivative with respect to $S'_m$ that appears in the numerator (\ref{num2}).
Again, we reiterate that  the second $W^{\rm gm}$  has to evaluated at $\delta_m=\delta_b$ as well as  its derivatives with respect to $\delta_m$.

The contribution from $N_c$ vanishes because it contains  a total derivative $\partial_n$
of a quantity that vanishes at $\delta_n=\delta_c$.

The term $N_a$ is immediately obtained using 
eqs.~(108) and (109) of MR1, and is given by
\bees
N_a&=& {\kappa\over \pi}
\left[ 
\sqrt{2\pi}{\delta_b\over \sqrt{S'_m}}{\rm e}^{-{(2\delta_b-\delta_m)^2\over 2S'_m}}\right.\nn\\
&&
-\sqrt{2\pi}{\delta_b \sqrt{S_m'}\over S_m}{\rm e}^{-{(2\delta_b-\delta_m)^2\over 2S'_m}}
\nn\\
&&\left. +\pi\frac{\d_b(\d_b-\d_m)}{S_m}
{\rm Erfc}\[\frac{2\d_b-\d_m}{\sqrt{2S'_m}}\]
\right]\nn\\
&&\times
\left.\pa_m\Pi^{\rm gm}(\d_m;\d_n;S_n-S_m)\right|_{\delta_m=\delta_b}\, .
\ees
The corresponding flux rate is given by
\begin{eqnarray}
{\cal F}^{(a)}_{\rm sph}(\delta_c, S_n|\delta_b, S_m)&=&
{\kappa\over 2}{\rm e}^{-{(\delta_b-\delta_c)^2\over 2(S_n- S_m)}}{\sqrt{S_m}\over (S_n-S_m)^{5/2}}\nn\\
&&\hspace{-3cm}\times \left(S_m-S_n+(\delta_b-\delta_c)^2\right)\,{\rm e}^{{\delta_b^2\over 2 S_m}}
{\rm Erfc}\left[{\delta_b\over \sqrt{2 S_m}}\right]
\,.
\label{Fa}
\end{eqnarray}

The term $N_b$ is given by
\bees\label{Nb}
N_b&=&\Pi^{\rm gm}(\d_0;\d_m;S'_m)\\
&&\times [
\Pi^{b1} (\d_b,S_m;\d_n,S_n) +\Pi^{b2} (\d_b,S_m;\d_n,S_n)]\, ,\nn
\ees
where 
\bees
&&\Pi^{b1} (\d_b,S_m;\d_n,S_n)\equiv 
\int_{-\infty}^{\delta_c}\hspace*{-2mm}
d\d_{m+1}\cdots d\delta_{n-1}\label{Nb1}\\
&&\times
\sum_{i=m+1}^{n-1}\Delta_{in}\partial_i\partial_n
W^{\rm gm}(\delta_b;  \ldots ,\delta_n; S_n-S_m)\, ,\nn
\ees
and
\bees
&&\Pi^{b2} (\d_b,S_m;\d_n,S_n)\equiv 
\int_{-\infty}^{\delta_c}\hspace*{-2mm}
d\d_{m+1}\cdots d\delta_{n-1}\label{Nb2}\\
&&\times
\frac{1}{2}\sum_{i,j=m+1}^{n-1}\Delta_{ij}\partial_i\partial_j
W^{\rm gm}(\delta_b;  \ldots ,\delta_n; S_n-S_m)\, .\nn
\ees
The computation of $\Pi^{b1}$ and $\Pi^{b2}$ is quite similar to the computation of the terms called
$\Pi^{\rm mem}$ and $\Pi^{\rm mem-mem}$ in MR1, and in the continuum limit $\eps\ra 0$ we get
\bees\label{Pib1}
&&\hspace*{-4mm}\Pi^{b1}(\d_b,S_m;\d_n,S_n)=\pa_n\lim_{\eps\ra 0}
\frac{1}{\eps}\int_{S_m}^{S_n}dS_i\,\nn\\
&&\hspace*{-4mm}\times\Delta(S_i,S_n)\Pi_{\eps}^{\rm gm}(\d_b;\d_c;S_i-S_m)
\Pi_{\eps}^{\rm gm}(\d_c;\d_n;S_n-S_i)\nn\\
&=&\frac{\kappa}{\pi} (\d_c-\d_b) \pa_n\biggl\{(\d_c-\d_n)
\int_{S_m}^{S_n}dS_i\nn\\
&&\times \frac{S_i}{S_n(S_i-S_m)^{3/2}(S_n-S_i)^{1/2}}\nn\\
&&\times
\exp\[-\frac{(\d_c-\d_b)^2}{2(S_i-S_m)}-\frac{(\d_c-\d_n)^2}{2(S_n-S_i)}\]\biggr\}\,
\label{finalb1}
\ees
and 
\bees
&&\hspace*{-4mm}\Pi^{b2}(\d_b,S_m;\d_n,S_n)=\lim_{\eps\ra 0}
\frac{1}{\eps^2}\int_{S_m}^{S_n}dS_i\int_{S_i}^{S_n}dS_j\,\nn\\
&&\hspace*{-4mm}\times \Delta(S_i,S_j)\Pi_{\eps}^{\rm gm}(\d_b;\d_c;S_i-S_m)\nn\\
&&\hspace*{-4mm}\times \Pi_{\eps}^{\rm gm}(\d_c;\d_c;S_j-S_i)
\Pi_{\eps}^{\rm gm}(\d_c;\d_n;S_n-S_j)\nn\\
&=&\frac{\kappa}{\pi\sqrt{2\pi}}\, (\delta_c-\d_b)(\delta_c-\delta_n)\nn\\
&&\hspace*{-4mm}\times
\int_{S_m}^{S_n}dS_i\,  \frac{S_i}{(S_i-S_m)^{3/2}}e^{-(\delta_c-\d_b)^2/[2(S_i-S_m)]}\nn\\
&&\hspace*{-4mm}\times\int_{S_i}^{S_n}dS_j
\frac{e^{-(\delta_c-\delta_n)^2/[2(S_n-S_j)]}}{S_j(S_j-S_i)^{1/2}
(S_n-S_j)^{3/2} }
\, .
\ees
This can be rewritten as a total derivative with respect to $\d_n$, as
\bees
&&\hspace*{-4mm}\Pi^{b2}(\d_b,S_m;\d_n,S_n)=
\frac{\kappa}{\pi\sqrt{2\pi}}\, (\delta_c-\d_b)\pa_n\nn\\
&&\hspace*{-4mm}\times
\int_{S_m}^{S_n}dS_i\,  \frac{S_i}{(S_i-S_m)^{3/2}}e^{-(\delta_c-\d_b)^2/[2(S_i-S_m)]}\nn\\
&&\hspace*{-4mm}\times\int_{S_i}^{S_n}dS_j
\frac{e^{-(\delta_c-\delta_n)^2/[2(S_n-S_j)]}}{S_j(S_j-S_i)^{1/2}
(S_n-S_j)^{1/2} }
\, .\label{finalb2}
\ees
The fact that both $\Pi^{b1}$ and $\Pi^{b2}$ can be written as a derivative with respect to $\d_n$ simplifies considerably the computation of the contribution of $N_b$ to the numerator (\ref{num2}), since we can integrate $\pa_n$ by parts, and then we only need to evaluate the integrals  in \eqs{finalb1}{finalb2} in $\d_n=\d_c$, which can be done analytically, as discussed in MR1. In particular the contribution to the numerator from $\Pi^{b1}$ vanishes because it is a derivative $\partial_n$ of a quantity that vanishes at $\delta_n=\delta_c$. Following the Appendix of \cite{ma}, 
the contribution to the two-barrier first crossing rate from  $N_b$ can be easily computed to be 
\begin{eqnarray}
\label{b}
&&\hspace*{-3mm}{\cal F}^{(b)}_{\rm sph}(\delta_c, S_n|\delta_b,S_m)=
-\frac{\pa}{\pa S_n}\Bigl[ 
\frac{\kappa (\delta_c-\d_b)}{\sqrt{2\pi S_n}}\nn \\
&&\times
\int_{S_m}^{S_n}dS_i\,  \frac{S^{1/2}_i}{(S_i-S_m)^{3/2}}
e^{-\frac{(\delta_c-\d_b)^2}{2(S_i-S_m)}}
\Bigr]\, .
\ees
The most complicated term is $N_d$. 
We get
\bees
N_d&=& {\kappa\over \pi}
\left[ 
\sqrt{2\pi}{\delta_b\over \sqrt{S'_m}}{\rm e}^{-{(2\delta_b-\delta_m)^2\over 2S'_m}}
{{\cal I}_2\over \pi}
\right.\nn\\
&&
-\left(\sqrt{2\pi}{\delta_b \sqrt{S_m'}}{\rm e}^{-{(2\delta_b-\delta_m)^2\over 2S'_m}}
\right.
\nn\\
&&\left.\left. -\pi{\d_b(\d_b-\d_m)}
{\rm Erfc}\[\frac{2\d_b-\d_m}{\sqrt{2S'_m}}\]
\right){{\cal I}_1\over \pi}
\right]\nn\\
&&+{\kappa\over \pi}\Pi^{\rm gm}(\delta_m; S'_m)
\partial_m \left.\left[ S_m {\cal I}_2-S_m^2{\cal I}_1\right]\right\vert_{\delta_m=\delta_b}\nn\\
&&+{\kappa\over \pi}\partial_m \Pi^{\rm gm}(\delta_m; S'_m)
 \left.\left[ S_m {\cal I}_2-S_m^2{\cal I}_1\right]\right\vert_{\delta_m=\delta_b}\,.
\ees
where
\begin{eqnarray}
{\cal I}_1&=&\left.
\partial_n\partial_m {\cal J}(\delta_m,\delta_n)\right\vert_{\delta_m=\delta_b}\,,\\
{\cal I}_2&=&
\int_{S_m}^{S_n}dS_j
\frac{(\delta_c-\delta_m)(\delta_c-\delta_n)}{ (S_j-S_m)^{3/2}(S_n-S_j)^{3/2}}\nn\\
&&\hspace*{-3mm}\times \exp\left\{-\frac{(\d_c-\d_m)^2}{2 (S_j-S_m)}
-\frac{(\d_c-\d_n)^2}{2 (S_n-S_j)}\right\} \nn\\
&=&\sqrt{2\pi\over S_n-S_m}\partial_n
{\rm e}^{-{(2\delta_c-\delta_m-\delta_n)^2}/ [2(S_n-S_m)]}\,, \\
{\cal J}&\equiv &\int_{S_m}^{S_n}dS_j
\frac{1}{S_j (S_j-S_m)^{1/2}(S_n-S_j)^{1/2}}\nn\\
&&\hspace*{-3mm}\times \exp\left\{-\frac{(\d_c-\d_m)^2}{2 (S_j-S_m)}
-\frac{(\d_c-\d_n)^2}{2 (S_n-S_j)}\right\}\, .
\end{eqnarray}
We are only interested in its value for $\d_n=\d_c$ as $N_d$ contains a total derivative with respect to
$\d_n$
\bees
{\cal J}(\d_m,\d_n=\d_c)
&=&\frac{\pi}{(S_mS_n)^{1/2}} e^{+(\d_c-\d_m)^2/(2 S_m)}\label{Idndc}\\
&&\times
{\rm Erfc}\[(\d_c-\d_m)\sqrt{\frac{S_n}{2S_m(S_n-S_m)}}\]\, .\nn
\ees
The corresponding flux rate is given by
\begin{eqnarray}
&&\hspace{-8mm}
{\cal F}^{(d)}_{\rm sph}(\delta_c, S_n|\delta_b, S_m)=
{\kappa\over \sqrt{2\pi}}{{\rm e}^{-{(\delta_c-\delta_b)^2\over 2(S_n- S_m)}}\over (S_n-S_m)^{5/2}}
\bigg\{ \nn\\
&&\hspace{-8mm}
\delta_b\left(S_m-S_n+(\delta_c-\delta_b)^2\right)
+(\delta_c-\delta_b)S_m\left(3-{(\delta_c-\delta_b)^2\over S_n-S_m}\right)\nn\\
&&\hspace{-8mm}
-{\sqrt{\pi S_m} \over S_n^2}
\left[
{\delta_b \over \sqrt{2\pi S_m}}+
{1\over 2}{\rm e}^{{\delta_b^2\over 2S_m}}{\rm Erfc}\left[{\delta_b\over \sqrt{2S_m}}\right]\right]\nn\\
&&
\times \Bigg[
\sqrt{2}S_n\left(-S_n^2+S_m\left(S_n+(\delta_c-\delta_b)^2\right)\right)\nn\\
&&\;\;\;\;
+\sqrt{\pi}{\rm e}^{{(\delta_c-\delta_b)^2 S_n\over 2 S_m(S_n-S_m)}}\sqrt{S_n\over S_m}
(S_n-S_m)^{5/2}(\delta_c-\delta_b) \nn\\
&&\qquad\times
{\rm Erfc}\left[\sqrt{{S_n\over S_m}}{\delta_c-\delta_b\over \sqrt{2(S_n-S_m)}}\right]
\Bigg]\nn\\
&&\hspace{-8mm}
+(\delta_c-\delta_b)
{\left(S_m^3-4S_m S_n^2+S_n^3+S_m^2(2S_n+\left(\delta_c-\delta_b)^2\right)\right)\over S_n(S_n-S_m)}
\nn\\
&&\hspace{-8mm}
-\sqrt{{\pi\over 2}}
{\rm e}^{{(\delta_c-\delta_b)^2 S_n\over 2 S_m(S_n-S_m)}}
{(S_n-S_m)^{5/2}\over S_m^{1/2} S_n^{3/2}}\left(S_m+(\delta_c-\delta_b)^2\right)\nn\\
&&\qquad\times
{\rm Erfc}\left[\sqrt{{S_n\over S_m}}{\delta_c-\delta_b\over \sqrt{2(S_n-S_m)}}\right]
\bigg\}\,.
\label{Fd}
\end{eqnarray}
The total two-barrier first-crossing rate is finally obtained by adding to the usual rate the corrections given by eqs.~(\ref{Fa}), (\ref{b}) and (\ref{Fd}):
\begin{eqnarray}
\hspace*{-2cm}
&&{\cal F}_{\rm sph}(\delta_c, S_n|\delta_b, S_m)=
{(\delta_c-\delta_b) {\rm e}^{-{(\delta_c-\delta_b)^2\over 2(S_n-S_m)}}\over 
\sqrt{2\pi} (S_n-S_m)^{3/2}}
\nn\\
&&+
{\cal F}^{(a)}_{\rm sph}(\delta_c, S_n|\delta_b, S_m)+
{\cal F}^{(b)}_{\rm sph}(\delta_c, S_n|\delta_b, S_m)\nn\\
&&+
{\cal F}^{(d)}_{\rm sph}(\delta_c, S_n|\delta_b, S_m)\,.
\end{eqnarray}
In the limit $S_m\ll S_n$, the total rate reduces to
\begin{eqnarray}
\hspace*{-2cm}
&&{\cal F}_{\rm sph}(\delta_c, S_n|\delta_b)=
{1-\kappa\over \sqrt{2\pi}}
{(\delta_c-\delta_b)  {\rm e}^{-{(\delta_c-\delta_b)^2\over 2S_n}}\over 
 S_n^{3/2}}\nn\\
&&+
{\kappa\over 2\sqrt{2\pi}}
{\delta_c-\delta_b\over S_n^{3/2}}\, \Gamma\left(0, {(\delta_c-\delta_b)^2\over 2 S_n}\right)
\nn\\
&&
-{\kappa\over \sqrt{2\pi}} {\delta_b\over S_n^{3/2}}
\left[1-{(\delta_c-\delta_b)^2\over S_n}\right]
  {\rm e}^{-{(\delta_c-\delta_b)^2\over 2S_n}}
 \,,
 \label{azz}
\end{eqnarray}
which reproduces the result of \cite{ma}.
In the case of a moving barrier, we expect that it is a  good approximation (\cite{Giocoli2007})
to simply replace the constant barrier $\delta_c$ with $\sqrt{a}\delta_c$,
and $\kappa$ with $a\kappa$ (\cite{MR2, ma}). If so, we obtain
\begin{eqnarray}
&&\hspace*{-0.5cm}
{\cal F}_{\rm mb}(\delta_c, S_n|\delta_b)=
{1-a\kappa\over \sqrt{2\pi}}
{\sqrt{a}(\delta_c-\delta_b) {\rm e}^{-{a(\delta_c-\delta_b)^2\over 2S_n}}\over 
 S_n^{3/2}}\nn\\
&&\hspace*{-0.5cm}+
a^{3/2}{\kappa\over 2\sqrt{2\pi}}
{(\delta_c-\delta_b)\over S_n^{3/2}}\, \Gamma\left(0, {a(\delta_c-\delta_b)^2\over 2 S_n}\right)
\nn\\
&&\hspace*{-0.5cm}
-a^{3/2}{\kappa\over \sqrt{2\pi}} {\delta_b\over S_n^{3/2}}
\left[1-{a(\delta_c-\delta_b)^2\over S_n}\right]
 {\rm e}^{-{a(\delta_c-\delta_b)^2\over 2S_n}}
 \,.
\end{eqnarray}
Instead, the case of the diffusive barrier is obtained from eq.~(\ref{azz}) by simply sending $S_n$ into
$S_n/a$ and $\kappa$ into $a\kappa$, that is
\begin{eqnarray}
&&\hspace*{-0.5cm}{\cal F}_{\rm dif}(\delta_c, S_n|\delta_b)=
{1-a\kappa\over \sqrt{2\pi}}a^{3/2}
{(\delta_c-\delta_b) {\rm e}^{-{a(\delta_c-\delta_b)^2\over 2S_n}}\over 
 S_n^{3/2}}\nn\\
&&\hspace*{-0.5cm}
+
{a^{5/2}\kappa\over 2\sqrt{2\pi}}
{\delta_c-\delta_b\over S_n^{3/2}}\, \Gamma\left(0, {a(\delta_c-\delta_b)^2\over 2 S_n}\right)
\nn\\
&&\hspace*{-0.5cm}
-{a^{5/2}\kappa\over \sqrt{2\pi}} {\delta_b\over S_n^{3/2}}
\left[1-{a(\delta_c-\delta_b)^2\over S_n}\right]
 {\rm e}^{-{a(\delta_c-\delta_b)^2\over 2S_n}}
 \,.
 \label{azzz}
\end{eqnarray}
We expect these results to be correct in the high mass limit.

\bibliographystyle{mn2e}

\label{lastpage}

\end{document}